Anomalous Silicate Dust Emission in the Type 1 LINER Nucleus of M81

Howard A. Smith

Harvard-Smithsonian Center for Astrophysics, 60 Garden Street, Cambridge, MA 02138

*hsmith@cfa.harvard.edu*

Aigen Li, M.P. Li, and M. Köhler

Department of Physics and Astronomy

University of Missouri, Columbia, MO 65211

M. L. N. Ashby, G. Fazio, J-S Huang, M. Marengo, Z. Wang, S. Willner,

Harvard-Smithsonian Center for Astrophysics, 60 Garden Street, Cambridge, MA 02138

A. Zezas [1]

Physics Department, University of Crete, 71003 Heraklion, Greece

L. Spinoglio

Istituto di Fisica dello Spazio Interplanetario, CNR, via Fosso del Cavaliere

100, I-00133 Rome, Italy

and

Y.L. Wu

Spitzer Science Center, California Institute of Technology, Pasadena CA 91101

[1]Also IESL, Foundation for Research and Technology-Helas, 7110 Heraklion, Greece; Harvard-Smithsonian Center for Astrophysics, 60 Garden Street, Cambridge, MA 02138

ABSTRACT

We report the detection and successful modeling of the unusual 9.7μm Si--O stretching silicate emission feature in the type 1 (i.e. face-on) LINER nucleus of M81. Using the Infrared Spectrograph (IRS) instrument on *Spitzer*, we determine the feature in the central 230 pc of M81 to be in strong emission, with a peak at ~10.5μm. This feature is strikingly different in character from the absorption feature of the galactic interstellar medium, and from the silicate absorption or weak emission features typical of galaxies with active star formation. We successfully model the high signal-to-noise ratio IRS spectra with porous silicate dust using laboratory-acquired mineral spectra. We find that the most probable fit uses micron-sized, porous grains of amorphous silicate and graphite. In addition to silicate dust, there is weak PAH emission present (particularly at 11.3μm, arising from the C--H out-of-plane bending vibration of relatively large PAHs of ~500--1000 C atoms) whose character reflects the low-excitation AGN environment, with some evidence that small PAHs of ~100--200 C atoms (responsible for the 7.7μm C--C stretching band) in the immediate vicinity of the nucleus have been preferentially destroyed.

Analysis of the infrared fine structure lines confirms the LINER character of the M81 nucleus. Four of the infrared $H_2$ rotational lines are detected and fit to an excitation temperature of T~800K. Spectral maps of the central 230pc in the [NeII] 12.8μm line, the $H_2$ 17μm line, and the 11.3μm PAH C--H bending feature reveal arc- or spiral-like structures extending from the core. We also report on epochal photometric and spectroscopic observations of M81, whose nuclear intensity varies in time across the spectrum due to what is thought to be inefficient, sub-Eddington accretion onto its central black hole. We find that, contrary to the implications of earlier photometry, the nucleus has not varied over a period of two years at these infrared wavelengths to a precision of about 1%.

1. INTRODUCTION

M81 is one of the nearest large spirals and a member of an interacting group of about 25 galaxies that includes M82. Its estimated distance is 3.63 ± 0.34 Mpc (Freedman, 1994), a value we use in all subsequent calculations. The nucleus of M81 has been classified as a LINER (Heckman 1980; Ho *et al.* 1997) based on a wide range of observations, with properties that mark it as being closely related to Seyferts. Ho et al. (1996) used HST and ground-based observations to probe the inner few parsecs, and concluded that the main source of excitation is photoionization from a non-stellar continuum, consonant with the conclusion that it is a genuine AGN with a likely accretion rate of ~$2 \times 10^{-5}$ $M_\odot yr^{-1}$. Its estimated black hole mass is $7 \times 10^7 M_\odot$ (Devereux et al. 2003; 2007), it has a nuclear X-ray (0.5-2.4keV) luminosity of $5 \times 10^6$ $L_\odot$ (Devereux and Shearer 2007), and a circumnuclear (64" radius) total IR luminosity of ~$4 \times 10^8$ $L_\odot$ (Perez-Gonzalez *et al.* 2006; Devereux *et al.* 1995). HST WFPC2 images of the nucleus in V-band find a single object less than 0.04" (0.7pc) in size, with no other nuclei or stellar components within about 4" (Devereux, Ford and Jacoby 1997); ground-based 10 μm data at ~arcsecond resolution are consistent with a point-like nucleus. Devereux, Ford and Jacoby (1997) and Davidge and Courteau (1999) studied the inner 1 kiloparsec (including the region whose *Spitzer* spectrum we discuss shortly) and its near-IR spectral energy distributions (SED) in J, H, K, and 2.26μm filters, concluding that thermal emission from hot dust can (as in other galaxies) contribute about 20% of the light at K-band within 0.5" of the nucleus. Davidge and Courteau interpret the hot dust as a sign of nuclear activity, and also conclude there is no population of extremely red stars in the central arcsecond. Although the M81 nucleus is relatively inactive, as is characteristic of LINERS, temporal variations have been reported across the spectrum due to what is thought to be inefficient, sub-Eddington accretion onto its central black hole. The nuclear emission varies by a factor of about 2 in visible light (Devereux et al. 2003) and near 10μm (Grossan et al. 2001) on timescales estimated at about 25 years.

Willner *et al.* (2004; hereafter Paper I) published the first IRAC observations of M81, including the nuclear region. The IRAC flux densities in a 3.5" square aperture centered on the nucleus (including starlight) were 109, 75, 65, and 56 mJy at 3.6, 4.5, 5.8, and 8 μm, respectively, with uncertainties of ±10%. After subtraction of the stellar component, they found a residual point-like source evident at 4.5μm and longer wavelengths, and perhaps at 3.6μm as well. Flux densities of the point-like source at the four IRAC wavelengths were 6, 21, 22, and 39 mJy, respectively, with the first being uncertain by more than a factor of 2. Willner *et al.* compared the 8 μm flux density with the reported, much stronger 10 μm flux density of 159 mJy observed in 1999 by Grossan *et al.*, (2001), and suggested based on this comparison that the nuclear source may have varied over the previous 4 years. They noted, however, that the spectral shape of the nuclear emission between 8 and 10 μm was not well known, although an ISO spectrum of the source in a 24" beam did show a steep rise (Rigopoulou et al. 1999). Since it would seem difficult to account for the differences in measured flux densities by spectral shape alone -- it would require a factor of two increase in the continuum level between 8 and 10 μm -- the authors suggested that the difference could instead be due to variability, thus supporting the idea that IRAC was seeing AGN activity. The IRAC images also revealed a filament or bar of material leading from a ringlike structure at ~1 kpc (1') to an inner structure and to the AGN. The prominence of the ring at 8μm suggests that much of its emission is from dust, probably mostly aromatic features. These inner features were suspected of being related to the structures seen in HST images and in the light of Hα (Devereux et al. 1997).

The possibility of time variations in the infrared flux from the M81 nucleus convinced us to begin a series of epochal observations of the source with IRAC Guaranteed Time; the possibility that an unusual spectral energy distribution (SED) might be responsible for the apparent discrepancy prompted us to examine the IRS spectra in the *Spitzer* archive. Photometric redshift estimates are powerful and increasingly common ways to estimate redshifts of faint, distant galaxies, and for galaxies out to $z \sim 2$ the 8-10μm SED can be sampled from IRS spectra (e.g., Huang & Faber, 2009). A steep and unusual SED in M81 (and, as we discuss below, in some other LINERs and quasars) will strongly influence the estimated photometric redshifts of any cosmological objects of a similar nature.

In Section 2 we review all the observations; section 3 presents an analysis of the peculiar SED in M81 and its spectral features, and compares it to the SEDs seen in other galaxies. Section 4 addresses the spectral lines in the nucleus. Section 5 models the SED as arising from silicate dust, and discusses the nature of that dust and possible explanations for its presence. Section 6 presents our spectral imaging data of the nuclear region; the conclusions are in Section 7.

## 2. OBSERVATIONS AND ANALYSIS

### 2.1 *IRAC, MIPS and IRS Observations*

After the original Willner *et al.* (2004; PID 1035) conclusions were published, we began a four-epoch campaign to look for possible variability of the M81 nucleus (Fazio, PID 121). This was motivated in part by Grossan *et al.* (2001), who reported from their ground-based N-band (10.79μm) observations compared to earlier observations from Rieke and Lebofsky (1978) that there was a factor of two flux variability, and that therefore the M81 nucleus had a substantial nonstellar component to its flux at 10μm. The Willner observations were taken on 2003 November 6, 2004 December 21, 2005 May 6, and 2005 October 24. These, and in addition the observations from SINGS on 2004 June 4, were reanalyzed using the IRACproc software package (Schuster, Marengo, and Patten 2006; e.g., Ashby *et al.* 2009). The combined, reanalyzed three-color IRAC image (3.6, 4.5, and 8.0 μm) spanning 1.7 years is shown in Figure 1a after subtraction of a stellar contribution as per Paper I; Figure 1b is a zoom into the nuclear region (inner 30") with its ring and filamentary arcs. Figure 1c, for comparison, is of the same nuclear region as seen with HST in Hα (Devereux, Ford and Jacoby, 1997), and shows the same arc-like structures. All of the features noted in Paper I are even more clearly seen in the deeper IRAC mosaic, in particular the central point-source. We employed a simple, iterative scale-and-subtract procedure to determine whether any changes in the apparent brightness of M81's nucleus could be detected in epochs subsequent to 2003 November. Specifically, we added a constant offset to the epochal mosaics to match the background in the 2003 November mosaic. We then determined the scale factors needed to produce residuals with a mean average of zero when the subsequent epoch mosaics were subtracted from the 2003 November mosaic.

This procedure was repeated for all epochs in all four IRAC channels. In some cases, small sub-pixel offsets were necessary to achieve proper spatial alignment.

We find that in all epochs, for all channels, we obtain scale factors that are consistent with unity to within $2\sigma$ or less. At 3.6, 4.5, 5.8, and 8.0μm, respectively, the scale factors derived are 0.991± 0.005, 0.991±0.006, 0.977±0.016, and 0.985±0.015. Given the errors attending this procedure, our results suggest that the variability of the M81 nucleus is less than 1% at 3.6 and 4.5μm, and less than 2% at 5.8 and 8.0μm. Given that these results are comparable to the 3% error in the IRAC absolute gain calibration and the 1% errors attending repeated point-source photometry according to v3.0 of the IRAC Data Handbook, we cannot rule out unknown systematic effects that would dominate our photometry at the level probed by our difference-image technique. Thus, we see no evidence for flux variability in the M81 nucleus in the IRAC bands.

The MIPS 24μm data were taken as part of the SINGS Legacy Survey program (Kennicutt et al. 2004). We downloaded these data from the Spitzer archive and reduced them independently. We used MIPS scan map observations from the PID 717 and PID 159 programs as generated by version S16.1 of the MIPS pipeline software. After first subtracting object-masked median stacked images on a by-AOR basis, the individual exposures were coadded in the standard way into three separate single-AOR mosaics with 2.5 arcsecond pixels. The flux within a 6x5 pixel region centered on the galaxy nucleus was measured in each of the three mosaics produced (two from PID 159, and one from PID 717). The result, aperture corrected per the MIPS guidelines, is a 24μm flux of 0.50 +- 0.05 Jy. We also used the post-BCD 70μm mosaic from PID 159, which we downloaded from the Spitzer archive. As before, pipeline version S16.1 was used. Within a square region 16" wide centered on the nucleus of M81 we measured a 70μm flux of 0.8 ± 0.2 Jy. [1]

The IRS data were obtained as part of the SINGS Legacy Survey program (Kennicutt, PID 159, 193), on 2004 April 15, and 2005 April 11 and 22, and retrieved from the *Spitzer* archive. The data were processed with pipeline S17, and then manually reduced using CUBISM (Smith *et al.* 2007), and with the SMART (Higdon *et al.* 2004) software packages used for the spectral analyses using Gaussian line fit routines. (SMART routines were also used to manually extract the spectra as a check, using a tapered column fit, with care taken to include the nuclear region and to exclude as much of the off-nuclear regions as possible. The results were consistent with CUBISM reductions.) The CUBISM extractions gave spectral orders matched in flux at the overlap wavelengths to better than about 8%, both for low-res and for high-res extractions; matching the continuum levels between low-res and hi-res spectra was a bit worse, with the high-res continuum systematically about 15% higher than low-res and as much as 20% higher at wavelengths longer than about 35μm. We refined the results by cross-referencing the flux values to the IRAC and MIPS24 photometry of the region. Figure 2 shows the full, composite-module spectrum of the nuclear region extending over 230pc; Table 2 lists the observed lines and their strengths. We used the SINGS MIPS images of M81 to extract the 24μm flux of the nuclear region and compare it with the value measured with IRS. Both give $F_{24} = 0.30 \pm .03$ Jy. The agreement gave us added confidence in assembling the full spectrum for the region from the

---

[1] We note that this figure is about 1% of the total 70μm flux of M81. At 24μm, according to Dale et al. 2007, our 0.5 Jy measurement is about 10% of the total 24μm emission from the galaxy of 5.09 +/- 0.20 Jy

different IRS modules.

## 3. THE MID-IR DUST EMISSION

### *3.1 The Unusual Silicate Emission Profile of the M81 Nucleus*

The continuum SED of the M81 nucleus (Figure 2) is striking: it shows a strong silicate emission bump with its flux increasing by a factor of two between 8 and 10μm. In fact, the continuum shape in M81 changes by exactly the amount needed to explain the apparent discrepancy between the 1999 ground-based photometry and the IRAC photometry, and thus obviates the need for time-variability as an explanation. Thus we find no evidence for time variability, either within the internal set of epochal IRAC infrared data reported in Section 2, or in comparison with the filter-corrected, ground-based observations.

The silicate emission profile of the M81 nucleus warrants careful attention. The spectral shape is not only steeply rising, but it peaks at wavelengths of ~10.5μm, longer than the 9.7μm peak usually seen in most of the galactic sources (see Figure 3a). Its width is also much broader than that of the galactic sources: while the silicate emission feature of the M81 nucleus has a FWHM of ~4.2μm, the 9.7μm silicate absorption feature of the local interstellar medium (ISM) along the line of sight toward Cyg OB2 #12 has a FWHM of only ~2.6μm (Whittet et al.1997); the ISM sightline toward the Galactic Center object Sgr A*, with a FWHM of ~1.8μm (Kemper, Vriend, & Tielens 2004), has an even narrower 9.7μm silicate absorption feature. Also shown in Figure 3a is the silicate absorption feature of IRAS08572+3915, an ultraluminous IR galaxy (ULIRG). Similar to that of the galactic sources, it peaks at 9.8μm and has a FWHM of ~2.2μm (Spoon et al. 2006). In Table 1 we list the peak wavelengths and widths of the "9.7μm" silicate emission or absorption features of these objects.

We will show in Section 5 that the redshifted, broadened "9.7μm" silicate emission feature of the M81 nucleus can be satisfactorily explained in terms of porous grains. This unusual kind of emission was already recognized in PG quasars by Hao et al. (2005), and Siebenmorgen et al.(2005); Sturm et al. (2005) refer to it as "inconsistent" with standard silicate ISM dust, and it was noted as "anomalous" and emphasized by Li, Shi, & Li (2008) in their discussion of the silicate emission profiles of type 1 AGNs using the examples of the quasar 3C 273 and the low-luminosity AGN NGC3998 (see Figure 3b). In the case of M81, we can spatially discriminate the nucleus from most extra-nuclear activities, something that is not possible with the IRS instrument for these other objects. For example, NGC 3998, at a distance of ~18Mpc, is five times farther away than M81, and the IRS short-low slit width corresponds to ~320pc at this distance. At the distance of M81 the slit width is ~65pc, and the integrated region whose SED is plotted in Figure 2 covers 230pc; for comparison, the possible bar in M81 extends to ~500pc (Elmegreen, Chromey & Johnson 1995), and the inner Lindblad resonance in M81 is estimated to be at ~1.2kpc (Westfahl 1998).

In order to facilitate the following discussion of the steeply rising SED of M81 in the 8-10 μm range, and to provide a more quantitative measure of its character as compared with that of other galaxies (Figure 4), we introduce the simple linear slope parameter $\gamma 810 \equiv [F_\nu(10\mu m) - F_\nu(8\mu m)/(2* F_\nu(9\mu m)]$ as a useful, first-order measure of the SED behavior in this range. Although this linear parameter cannot separate steep, monotonically rising (i.e., red) SEDs from rising SEDs caused by spectral bumps due to silicate emission, nor can it accurately quantify the wavelength of the peak or correct for the redshifts of distant galaxies in automated spectral searches, it is a simple way to discuss the first-order slope of nuclei, and it allows users to scan hundreds of IRS spectra in the archive (and in models) to identify ones suitable for more careful analysis. For most cases, $\gamma 810$ provides an estimate of the strength of the silicate feature: positive values signal probable silicate emission, while negative or very negative values -- the most commonly seen values as we will discuss next – signal silicate absorption. In M81, $\gamma 810$ $=0.37 \pm 0.04$, a remarkable large, positive value. We examine next the shape of the mid-infrared SED in a variety of galaxy types to see whether the M81 nucleus belongs to a wider family.

### 3.2 *The Mid-IR dust Emission Character of Other Galaxies*

Hundred of galaxies have now been studied with *Spitzer* IRS. Their spectral shapes in the 10μm region reflect a wide range of dust conditions, but the many papers analyzing the spectra of one or another of the morphological types typically report a uniformity of shape within each class. Figure 4 displays the scaled IRS 10μm silicate band spectra of four galaxies – Mrk 231, M82, NGC4151, and NGC4472 -- whose SEDs are typical for their classes (discussed below), along with that of the M81 nucleus (many of the cited papers explicitly present an average spectrum for a class). Starburst galaxies and ULIRGS produce some dust emission in their starbursts, but strong absorption by a more extended, cold component in these galaxies results in masking the emission, producing negative (or occasionally small positive) values of $\gamma 810$ (e.g., Smith *et al.*, 2007; Weedman *et al.,* 2006; Armus *et al.,* 2007; Sirocky *et al.,* 2008). AGN like Mrk 231 span a more complex range of SEDs and $\gamma 810$ values, a measure of the important diagnostic value of this spectral interval, with some showing silicate absorptions but others with small but positive 8-10μm slopes (Lutz *et al.*, 1998; Sturm *et al.*, 2000; Weedman *et al.*, 2006). Seyfert galaxies (e.g., NGC 4151) generally do not show evidence for strong silicate features in emission or in absorption in either Type 1s or Type 2s (Tommasin et al., 2008, 2010; Buchanan *et al.*, 2006; Wu et al., 2009), but a few do have strong absorption features, and a few others have steep, linearly rising SEDs but without evidence for a distinct feature (Tommasin et al, 2009).

PG quasars, in contrast to AGN and starburst galaxies, show strong dust emission resembling that of M81 (e.g., Haas, 2001; Hao *et al.*, 2005; Sturm *et al.*, 2005; Netzer et al., 2007). In some cases the 8-10μm slopes are even more positive, with a maximum $\gamma 810 \sim +0.45$ and an average value of $<\gamma 810> = +0.25$. Other quasars have mid-infrared spectra with small slopes and more closely resemble less-luminous AGN, those dominated by signs of star formation, than these PG or BAL quasars (Haas et al., 2005, 2008; Polletta *et al.*, 2006, 2008). The mid-IR emission of elliptical galaxies (e.g., NGC 4472; Figure 4) were studied by Bregman, Temi, and Bregman (2006), who report evidence of modest silicate emission that they argue is the result of silicate grains formed in the atmospheres giants with strong AGB winds. The average value of $\gamma 810$ for these galaxies is 0.04; the slope of the emission is much shallower than in M81, and the spectra peak about at 10μm, about 0.5 μm shorter than in M81.

LINERS, a class of which M81 is a member (S1.5; Ho *et al.* 1997), show a range of properties as explored by Sturm et al. (2005; 2006a,b), with silicate absorptions that range from weak or very weak to strong absorption. M81 is one of the exceptions, having strong silicate emission. Another such case is the faint type 1 source NGC 3998 (Sturm *et al.*, 2005; 2006a) which shows strong silicate emission ($\gamma 810 = 0.35 \pm 0.08$) very similar to that of the M81 nucleus. Another case is NGC 7213 (included in Wu et al., 2009), also a Seyfert, whose SED is strongly increasing ($\gamma 810$ = +0.53; see also Wu *et al.,* 2010). Only a limited number of IRS spectra of type 1 LINERs are available, but these LINERS, unlike ULIRGs, starbursts, or normal AGN, seem to show a more complex range of 10μm spectral features, sometimes with silicate in strong absorption, and at other times with strong silicate emission. LINERs span a complex blend of physical properties, often with confused identifications (e.g., Ho, Filippenko, and Sargent, 1993), and/or are thought by many to be a link between starbursts and AGN phases of evolution (e.g., Alsonso-Herroro et al., 2000) or to be helpful in challenging the strictly geometrical interpretation of type 2 AGN. Thus it may be natural that they also manifest a complex variety of characteristics at these infrared wavelengths where the dust features show a wide range of possible characteristics.

## 4 GAS SPECTRAL LINE ANALYSES

Eighteen spectral features – PAHs (poly-aromatic hydrocarbons), $H_2$ lines, and atomic fine structure lines – are detected in the low and/or the high resolution spectra of the M81 nucleus (Table 2). The table lists the line strengths originating in a region 13.0"x13.0" centered on the nucleus. The size of the region was set by the fact that it was observed in all six of the IRS spectral bands; using the capabilities of CUBISM (Smith *et al* 2007), we were able to extract line strengths from matched regions.

*4.1 $H_2$ Lines*

Three of the v=0-0 pure rotational $H_2$ lines are seen with high signal-to-noise ratios in the high resolution scans; although the low resolution mode readily detects the 0-0 S(1) line at 17.0μm, it does not detect the weaker 0-0 S(2) line at 12.28μm. The 0-0 S(5) line at 6.910 μm (and a possible [ArII] $^2P_{1/2} - ^2P_{3/2}$ line at 6.985μm) are not adequately resolved by IRS, and the feature there is only weakly detected; the fundamental 0-0 S(0) line at 28μm also has only an upper limit. Figure 6 is an energy diagnostic diagram of these lines. The normalized upper level population is directly calculated from measured flux and a distance of 3.63Mpc in the standard way (e.g., Tommasin *et al.* 2008; 2009). When compared with the energy of the upper level, the reciprocal slope of the plotted line gives an effective excitation temperature for the gas. A formal regression fit gives a temperature of ~825K (or ~515K if the S(5) lined measurement is treated as a limit); pair-wise limits from the higher signal-to-noise lines, range from 270K to over 1000K. If the gas were at 500K or more, it represents significantly warmer gas than typical values of ~200K – 300K obtained for most Seyferts (e.g., Tommasin et al. 2008; 2009). Further observations can refine this value. If it is indeed high, it suggests that the gas in regions around the nucleus is warmer than the average. Assuming all of the $H_2$ is at a minimum temperature of 270K and the ortho-para ratio is 3:1, the partition function can be calculated and the total mass in hot $H_2$ in the inner 230pc is ~$3\times10^5$ $M_\odot$. For temperatures of 515K (825K), the corresponding mass drops to ~6000 $M_\odot$ (~1500 $M_\odot$). With only these pure rotational lines it is not possible to determine the ultimate cause of the excitation, but shocks and/or ultraviolet radiation are common origins, and are expected to be present here.

*4.2 Fine Structure Lines*

The infrared fine structure lines of five elements in ten transitions are detected in the central 230 pc of M81 (Table 2; Figures 3, 5). There is strong emission in the NeII and NeIII lines, and the weak evidence for two NeV lines is most probably consistent with upper limits. In addition to neon, the high resolution mode of IRS measures the [SIII] 18.7 and 33.5μm lines and the [SIV] 10.5μm line, the [OIV] and [FeII] features at 25.9 μm, and the [Si II] 34.8 μm line. The central kiloparsec of M81 has been studied in detail in its optical lines (e.g., Devereux, Ford and Jacoby 1997; Devereux *et al.* 2003, Ho, Filippenko and Sargent 1996). The nucleus harbors a modestly active nucleus, and is clearly a LINER in its excitation parameters; most photoionization derives from a central UV source. The IRS spectra do not alter the current understanding of M81, but do lend some additional perspective. The lower ionization-state fine structure lines of neon, from [NeII] and [NeIII], can be used as a measure of star formation activity, at least in star forming regions. In the M81 nuclear region, the measured ratio [NeIII]15.6μm/ [NeII]12.8μm is 0.66, in the same range as starbursts (e.g., Thornley *et al* 2000); the ratio of the sum of the line fluxes, $2.4 \times 10^5$ $L_\odot$, to the luminosity value of region $L_{IR} = 4 \times 10^8$ $L_\odot$ (Perez-Gonzalez *et al* 2006), also conforms with starbursts (Ho and Keto, 2007). However, hydrogen recombination lines are predicted from starbursts (Ho and Keto, 20007) but are not seen in our data. Combined with evidence from the higher excitation lines, the overall conclusion is that pure galactic star formation is *not* primarily responsible for the fine structure excitation in the M81 nucleus, and there must be some excitation components from other sources, placing the M81 LINER nucleus in the anticipated class where typical, modest galactic starburst activity is found (e.g., Tommasin *et al* 2008; 2010).

The [OIV] $^3P_{3/2}$ - $^3P_{1/2}$ line at 25.89μm (ionization potential 54.9eV) is typically seen in LINERs (e.g., Tommasin *op cit;* Sturm *et al*. (2006), and is clearly detected here. We find the ratio of the of [OIV 25.9μm] / [NeII 12.8μm] lines to be 0.10 ± 0.01, comfortably in the range of 0.03 – 0.3 that Sturm et al. (2006) report for LINERS (and see their Figure 2). Other line ratios -- [FeII]/[OIV], [OIV]/[NeII], [FeII]/[NeII] and [FeII]/[SiII], and [SIV]/[SIII 18.7μm] – also confirm the LINER or transition LINER nature of M81 (Sturm *et al* 2006; Rupke et al. 2007), and place the M81 nucleus in the borderline AGN-starburst range line, near the galaxies with densities of ~3x10$^4$ cm$^{-3}$ and ionization parameters log U ~ -1.5 (U is the ionizing photon flux per unit area per hydrogen density; Spinoglio *et al.1995;* Tommasin *et al* 2008; 2010). The ratio we find for the [NeV 24.3μm] / [NeII 12.8μm] is ≤ 0.03, and the [NeV 14.3μm] / [NeII 12.8μm] is ≤ 0.02, consistent with the Seyfert and LINER results of Tommasin et al. (2010) who find the AGN-starburst transition galaxies have a [NeV 14.3μm] / [NeII 12.8μm] ≈ 0.08, as in the case of the type 2 LINER NGC 4922 (and see their Figure 3), with starbursts having lower values. The IRS high-resolution spectrometer has a resolving power of about 500 km-sec$^{-1}$; the FWHM of all of the observed fine structure lines are not quite resolved at ≈550 km-sec$^{-1}$. Although some of the optical fine structure lines in LINERS have widths over 2500 km-sec$^{-1}$ (Ho et al., 1996), those results are for the inner narrow-line region as measured in a 0.3" beam. Finally, as noted in Section 3.6, the Sturm *et al.(2006)* sample find that transition LINERs have continua with very small absorptions in their mid-IR SEDs, distinctly different from the strong M81 emission SED.

*4.3 PAH Emission Features*

Three infrared PAH features are clearly seen in the M81 nucleus: 11.3μm, 12.7μm, and 16.5μm (see Figures 2, 5). Table 2 lists the flux in each feature. The fluxes were determined by first subtracting any atomic lines, and then measuring the equivalent width of the observed features, with their continua obtained by interpolating at the features' centers. The 6.3μm feature is only weakly seen above the stellar and dust continuum, its strength depending on those fits; the PAH components of the 17μm complex besides the 16.5μm feature (Smith *et al*. 2007) are not seen, with an upper limit to the flux of $5 \times 10^{-22}$ watts-cm$^{-2}$. Significantly, there is also no clear sign of the 7.7μm feature, to a level of less than $1.2 \times 10^{-20}$ watts/cm$^2$, approximately 15% of the 11.3μm feature. The 11.3μm /7.7μm band ratio, even with very generous estimates, is ~5.7 – much larger than the typical value of ~0.3 of the galactic interstellar regions (see Figures 16, 18 of Draine and Li, 2001) starburst galaxies (e.g., see Bernard-Salas et al., 2009; Galliano et al., 2008), and some LINERS (e.g., Sturm et al., 2006). (One difficulty in ascertaining the strength of this feature in M81 is the strong and rapidly declining continuum from 5μm to 9μm arising from the spectral tail of the emission from photospheres of evolved stars, on which any putative feature is superimposed.) The 7.7μm feature is mainly emitted by PAH ions of ~100--200 C atoms (see Figures 6, 7 of Draine & Li 2007), and is notably absent in the vicinity of AGNs, perhaps because intense UV radiation preferentially destroys small PAHs while the relatively large PAHs with ~500-1000 C atoms – the ones that are responsible for the 11.3μm feature - survive (see Li 2007, and Figures 6, 7 of Draine & Li 2007). The hint of PAH at 6.3μm may suggest the presence of the 6.3μm C--C stretching feature, but its detection is far less secure than the much stronger 11.3μm feature, in part because of the difficulty in subtracting the strong underlying 5--9μm continuum.

Rigopoulou *et al. (*1999), Weedman *et al.* (2006), Spoon et al. (2006), and others used the relative strength of the 7.7 μm feature to infer AGN activity, and their several diagnostic schemes are all consistent the M81 nucleus being identified as a LINER. Sturm et al. (2006) further reported that the weak LINERs (both Types 1 and 2) have very weak or absent PAH emission in the 5-10μm range, but are bright at 11.3μm, unlike the bright infrared LINERs which look like starburst galaxies. O'Halloran, Satyapal and Dudik (2006) note that supernova activity can also destroy PAH molecules. Supernovae also enhance the amount of Fe in the ISM. Those authors use the [FeII]/[NeII] ratio as a measure of such activity, and find a linear relationship to the PAH 7.7μm feature in the sense that a low ratio correlates with strong PAH emission. Our very small value for the ratio, 0.09, implies in their scheme a very strong PAH feature, and therefore its absence cannot be explained in terms of strong supernova activity.

## 5. THE NATURE OF THE DUST IN THE M81 NUCLEUS

### 5.1 *Modeling the M81 Nucleus Silicate Emission*

It has been known for over 35 years that silicate grains produce spectral features in Seyfert galaxies at 10μm (e.g., see Rieke & Low 1975, Kleinmann, Gillett, & Wright 1976). The detection of silicates in the galactic ISM had been reported several years earlier, first in emission in the Trapezium region of the Orion Nebula (Stein & Gillett 1969), then in absorption toward the Galactic center (Hackwell, Gehrz, & Woolf 1970), and shortly thereafter toward the Becklin-Neugebauer object and Kleinmann-Low Nebula (Gillett & Forrest 1973). The numerous IRS studies of the many galaxy types and models discussed in Section 3 typically invoked silicate grains (e.g., see Hao et al. 2007; Levenson et al. 2007; Spoon et al. 2007; Thompson et al. 2009). The presence of some form of silicate dust in the M81 nucleus is clearly evidenced by the characteristic peaks around 10 and 18μm in its mid-IR emission spectrum (see Figure 2 and Section 3.1), attributed respectively to the Si—O stretching and O--Si--O bending modes of amorphous silicates. As discussed in Section 3.1, the 9.7μm silicate emission feature of the M81 nucleus is unusual in the sense that, compared to that of the galactic sources, it is significantly broadened and redshifted. Here we discuss fitting the silicate emission feature of the M81 nucleus in order to understand the nature and condition of the dust in the M81 nucleus.

The typical approach to modeling silicate features in galaxies has been to use the dielectric functions of "astronomical silicates" (Draine & Lee 1984; Laor and Draine, 1984). Siebenmorgen et al. (2005), Fritz, Franceschini & Hatziminaoglou (2006), Groves et al.(2006, 2008), Thompson et al. (2009), and Nikutta, Elitzur, & Lacy (2009) provided good fits to the mid-IR data of various galaxies using a combination of dust temperatures and geometries, including dust disks, torii, or compact, clumpy circumnuclear rings at various viewing angles. Modeling of the IR dust emission of AGNs has also been undertaken by numerous groups considering a hot, dusty torus seen at various angles. Schweitzer et al. (2006, 2008) successfully modeled a set of 23 quasars observed with IRS using silicate dust. These models predicted silicate in emission for type 1 sources, with details that depend on the nature of the dust as well as on the physical parameters of the circum-AGN region, but none of the models came close to predicting the silicate emission feature seen in M81.

We reiterate that M81 is not entirely unique in its having a 9.7μm silicate emission feature that is broadened and redshifted (e.g., see Figure 3b). Sturm *et al.* (2005) have already noticed that some PG quasars also have their 9.7μm silicate emission features broadened and redshifted in a similar way (see also Netzer *et al.*, 2007). They concluded that the quasar dust they saw is "inconsistent" with standard dust. In the case of NGC3998, Sturm *et al.* (2006) explicitly noted that the dust emission appears not to be due to "standard silicate ISM dust." Li, Shi, & Li (2008) suggested that the "peculiar" silicate emission feature in these quasars is consistent with that from dust that is highly porous; they estimate that the porosity P (the volume fraction of vacuum in a fluffy grain) might be as large as P>0.6. Markwick-Kemper et al. (2007) observed dust emission in the PG quasar PG2112+059. They found that the dust emission profile is also somewhat similar to that of M81. They were able to fit that spectrum to emission from a mixture of crystalline silicates (about 5% by weight), amorphous silicates (about 50%), corundum (about 38%), and periclase (about 6%).

We consider two dust populations, warm dust and cold dust, in fitting the M81 emission. The dust is modeled as a porous composite mixture of amorphous silicate, amorphous carbon and vacuum (see Li et al. 2008). Assuming the dust torus to be optically thin in the IR, the flux emitted by the dust is

$$F_\lambda = \frac{1}{d^2} \left[ A\, B_\lambda(T_\star) + m_{\rm w} \kappa_{\rm abs}(\lambda) B_\lambda(T_{\rm w}) + m_{\rm c} \kappa_{\rm abs}(\lambda) B_\lambda(T_{\rm c}) \right] \quad, \tag{1}$$

where d≈3.63Mpc is the distance of M81, $\kappa_{\rm abs}(\lambda)$ is the dust mass absorption coefficient, $m_{\rm w}$ and $m_c$ are the masses of the warm and cold dust components, respectively, $T_w$ ($T_c$) is the temperature of the warm (cold) dust component, $B_\lambda$ (T) is the Planck function of temperature T at wavelength $\lambda$, and A$\approx$1.96 $\times$ 10$^{32}$ cm$^2$, a constant, is invoked to account for the continuum emission at $\lambda <$ 8 $\mu$m which is approximated by the photospheric emission from M0III stars ($T_* \approx$ 3300 K).

The mass absorption coefficient of the composite silicate-carbon mixture $\kappa_{abs}(\lambda)$ is calculated using Mie theory together with the Bruggeman effective medium theory (Bohren & Huffman 1983). The dust size a and porosity P are treated as free parameters. The mass ratio of amorphous carbon to amorphous silicate is taken to be $m_{carb}/m_{sil}$ = 0.7, as estimated from the cosmic abundance constraints (see Li & Lunine 2003). For the dielectric functions of amorphous silicate and amorphous carbon, we adopt those experimentally determined, by Dorschner et al. (1995) for amorphous olivine MgFeSiO$_4$ and by Rouleau & Martin (1991) for amorphous carbon.

Figure 7 shows that the 5 - 35$\mu$m IRS emission spectrum can be best fit with two components of porous dust with a $\approx$5$\mu$m and P$\approx$ 0.5: warm dust with $T_w \approx$275K and $m_w \approx$0.034 M$_\odot$, and cold dust with $T_c \approx$ 40K and $m_c \approx$1500 M$_\odot$. The model fit to the entire ~5--35$\mu$m emission is excellent, except that (1) the model over-predicts <12% at the red wing (~19-24$\mu$m) of the "18$\mu$m" O--Si--O band [2], and (2) the model over-predicts <40% at the blue wing (~8--9.5$\mu$m) of the ``9.7$\mu$m'' Si--O band. The former over-prediction is probably due to the lack of knowledge of the exact optical properties of silicate dust at the 18$\mu$m band. Silicates of different

---

[2] The "18$\mu$m" O--Si--O band of the M81 nucleus actually peaks at ~17.2$\mu$m

compositions exhibit very different optical properties at this 18μm band, both in peak wavelength and in strength (e.g., see Dorschner et al.,1995). The latter over-prediction is probably due to the lack of knowledge of the actual stellar continuum emission (see Figure 7a).

The best-fit model, with a ≈ 5μm, requires the dust to be much larger than typical interstellar dust (for which the mean size is ~ 0.1μm). This is not surprising: in the dense circumnuclear region around AGNs, dust is expected to grow to large sizes via coagulation (e.g., see Maiolino, Marconi, & Oliva 2001). As illustrated in Figure 7b, for grains smaller than ~5μm their "9.7μm" silicate emission bands peak at too short a wavelength and their widths are also too narrow to be comparable with that observed in the M81 nucleus. For grains much larger than a=5μm, their "9.7μm" silicate emission features become too shallow and their peaks redshift too much to fit the data. We also note that a porous structure is expected for dust formed through coagulational growth, as demonstrated both theoretically (Cameron & Schneck 1965) and experimentally (Blum & Wurm 2008). This also explains why compact dust is less favored: it is unreasonable for dust to grow up to size a = 3.2μm while still retaining its compact structure (e.g. see Blum & Wurm 2008), although the model consisting of pure, compact silicate dust (with P=0 and a ≈3.2μm) is able to fit the observed mid-IR dust emission reasonably well (see Figure 8). Finally, we note that models with $0.5 \leq P \leq 0.7$ also work reasonably well (see Figure 9), with the temperatures of the warm component varying by ~10% and the masses of the warm component varying by ~15% (see Table 3). The mass of the cold component is less well constrained, as it is very sensitive to the adopted dust temperature and to the strength of the 18μm O--Si--O band of the particular adopted silicate material.

All of our above modeling uses the dielectric functions of amorphous olivine $MgFeSiO_4$ as experimentally measured by Dorschner et al.(1995). Although the exact stoichiometric composition of cosmic silicate dust is unknown, the cosmic abundance considerations suggest that $MgFeSiO_4$ appears to be a reasonable approximation (see Draine 2003). The widely-adopted dielectric functions of "astronomical" silicates (Draine & Lee 1984) were synthesized based on a combination of astronomical observations and laboratory measurements. They do not represent any real silicate materials. More specifically, the 8--13μm dielectric functions of "astronomical" silicates were derived based on the 9.7μm silicate emission profile of the Trapezium star-forming region (Draine & Lee 1984; see also Henning, 2010, Figure 12). To be complete, and for comparisons, we have also tried to fit the mid-IR silicate emission of the M81 nucleus using the dielectric functions of canonical "astronomical" silicates. As shown in Figure 10 for the best-fitting porous dust model and compact dust model, the results are not as good as that based on the experimentally measured dielectric functions of amorphous $MgFeSiO_4$ (Dorschner et al.1995). Finally, we have also tried a combination of the dust composition and features used by Markwick-Kemper et al.(2007) but were likewise unsuccessful (F. Markwick-Kemper, private communication).

If porous silicate dust grains are not unusual, it is reasonable to suspect that in some type 2 AGN, where the nucleus is seen through an absorbing layer of dust, the silicate absorption feature will also manifest the shape characteristic of these kind of grains. Roche et al. (1983) noted that in the case of Mrk231 the conventional 9.7μm silicate absorption feature seemed to be shifted to longer wavelengths. Inspection of the higher signal-to-noise archival IRS spectra of the nuclear region of Mrk231 (including PIDs 21110272 and 34294016; see also the reduction in Veilleux *et al* 2009) indeed confirms that the rest-frame absorption in Mrk231 has its minimum at ~10.2μm , consistent with there being cool, porous silicate dust in this object.   In a forthcoming paper (Köhler et al., in prep) we apply our porous dust models to the handful of known LINERS and galaxies, mentioned previously, that have similar silicate emission features peaking longward of the 9.7μm peak.

*5.2 Possible Origins of the M81 Dust*

Assuming that porous amorphous silicate and carbon dust is indeed present in M81, what would be the physical process responsible for producing it?  The question has broader implications because of the renewed interest in the mid- and far-IR emission of galaxies -- AGNs and ULIRGs in particular -- the nature of the dust that shapes the SEDs,  and not least the production of dust in very high-redshift galaxies that may not have had enough time to produce dust via AGB stars. AGB stars are an important source for dust grains, and are the primary source for PAHs and dust in galaxies (e.g., see Galliano, Dwek & Chanial 2008).   Bregman, Temi & Bregman (2006) identified the modest 10μm silicate signature they saw in ellipticals (with $\gamma 810$ ~0.04), with silicate grains produced in AGB star winds, as based on the models of Piovan et al. (2003) whose models then allow an estimate of the age of an elliptical galaxy.   AGB wind-produced dust could also be the primary constituent in M81.  Jura & Kleinmann (1989) estimated a total mass loss rate of  ~3-6x$10^{-4}$ $M_\odot$ yr$^{-1}$ from the AGB stars in the solar neighborhood.  In a densely populated area like the nucleus of M81, we might expect an even higher mass loss yield, considering that the steeply decreasing (blue) continuum at the short 5--8μm IRS band is probably the result of old stellar photospheres, with the steepness of the decline mostly arising from stars older than about 3 Gyr. In the central ~230pc of M81 probed by our observations, and for a standard dust to gas mass ratio of ~0.01, our mass estimate of 0.034 M$_\odot$ corresponds to a dust mass loss in excess of $\dot{M}_d$ ~$10^{-8}$ M$_\odot$ yr$^{-1}$.

Supernovae will also produce dust (e.g., Dwek, Galliano and Jones, 2007, 2009a,b; Galliano, Dwek and Chanial, 2008) . The observed amounts of dust produced by supernovae, however, seem to too small, perhaps by a factor of ten or more depending on the stellar initial mass functions (IMFs) and model details, in part because SN also destroy dust in their shocks (Dwek Galliano and Jones, 2007; 2009a,b) Observations confirm (e.g., Sugerman et al. 2006) that the amount of dust produced per supernova is small; for example, the Type II SN 1987a made only about $8 \times 10^{-4}$ M$_\odot$ of dust (Ercolano et al. 2007). Dust production by accretion in the ISM offers another option. Draine et al. (2007) argue that in both local galaxies and those at high-z, ISM grain re-growth must be the primary source of silicate dust because grain destruction is too rapid in the SN shocks (Draine and Salpeter, 1979).

Elvis, Marengo, and Karovska (2002) proposed the production of dust directly in the outflows of quasar winds. Since this initial paper there has been considerable interest in testing the idea. Although the black hole in M81 is currently quiescent, it may have had some type of outburst and/or wind in the past. The Spitzer IRS SEDs might be just such a quantitative handle on hot, emitting silicate dust. The above authors argued that the dust composition depends on the composition of the initial broad emission line clouds (BELCs) that the AGN processes. Since the BELC composition is not well understood, the grains we measure are at least consistent with it, and at best can be used to work backwards to study the metallicity of the BELCs. The amount of dust that can be produced by an AGN is estimated to be roughly $\dot{M} \sim L/L_{48}$ $M_\odot$-yr$^{-1}$, where L is the AGN luminosity and $L_{48}$ the luminosity in units of $10^{48}$ ergs-sec$^{-1}$. If we estimate the (unknown) prior activity of M81 to have a lifetime average of about 1% of the Eddington limit, then $L \sim (4\pi cGM_{BH}/\kappa)\, 0.01 = 5 \times 10^{42}$ ergs-sec$^{-1}$, and $\dot{M} \sim 5\times 10^{-6}$ $M_\odot$-yr$^{-1}$, where G is the gravitational constant, $M_{BH}$ is the black hole mass, and $\kappa$ is the opacity, so that the net amount of warm dust we observe can be assembled in a few hundred thousand years. Elvis, Marengo and Karovska (2001) also effectively address whether the dust, once made, can survive. Their conclusion is that it can, at least to within the general uncertainties of this plausibility estimate. As in the case of supernova production, a range of grain sizes would be produced but selectively depleted of small grains by radiation and shocks. A final key hypothesis in the Elvis, Marengo and Karovska scheme to have dust formation in the AGN wind is that such an outflow has a bipolar structure. The putative 14° inclination of the M81 AGN rotating disk (Devereux et al.,2003), however, is small enough that it would be very difficult to detect a flow in our data spatially if it were normal to the disk.

## 6 CIRCUMNUCLEAR STRUCTURE

The IRAC image of the nuclear region (Figure 1b), especially as seen at 8μm (Band 4), shows a bright central region with two clearly demarcated spiral "miniarms" emanating, one in the northeast originating at PA~30° and arcing clockwise by about 90° (the "northern" miniarm), and another originating at PA~ 170º and arcing clockwise over about 150º (the "southern" miniarm), apparently overlapping with the northern miniarm (1 arcsecond corresponds to 17.6pc). At the origin of the southern miniarm is a short stub of bright emission seen out to about 90 pc from the nucleus. Extensions to the nucleus are also seen in at radio wavelengths. At 8.4GHz, M81 has a small (about 700AU) one-sided relativistic radio jet extending to the northeast at PA~ 55º (Bietenholz, Bartel & Rupen, 2000; Markoff *et al.* 2008). This small jet shows variability in its length and shape over timescales of a year, including a few epochs in which small elongation is seen at the opposite side of the nucleus, but the major axis of the extended emission does not change direction significantly (Bietenholz, Bartel & Rupen *et al* 2000). The third notable feature of the 8μm image is a distinct loop of material inside the northern miniarm whose maximum extent occurs about 200pc from the center at PA~ 20º; in addition there are several other wisps visible. We find no evidence in the IRAC images of the bar-like feature in the nucleus that was reported at shorter infrared wavelengths (Elmegreen, Chromey & Johnson 1995).

The IRS spectrometer obtained small maps of the nuclear region – each of the slits covered a somewhat different area. All of them covered the inner 13.0"x13.0" zone, however, from which the line fluxes are extracted; we obtained them using the CUBISM extraction routines. In particular, our earlier discussion of the dust properties is limited to this inner 230pc of M81 so that the long wavelength IRS results can be compared to the short wavelength ones, and so that both 10 and 18μm silicate features can be reasonably modeled together even though at long wavelengths the pixel size is at almost three times bigger. It is possible, however, to extract other spatial information from larger maps, though it is restricted to those features seen in each particular observing mode. Figure 11 contains a series of these larger-area, single feature extractions.

The nuclear arcs seen at 8μm (Figure 1b) are clearly apparent in the 11.3μm PAH map (Figure 11a), which shows a very close correspondence to the IRAC image (the IRS map does not extend beyond the region shown to cover the full IRAC image displayed). Devereux et al. (1995) obtained an image of the nuclear region in Hα emission (Figure 1c) that is also remarkably similar to the IRAC image and the IRS 11.3μm image; in particular, it also clearly shows the two miniarms and the loop. Possible contributions to the 8μm IRAC image by the 7.7μm PAH feature are modest since this feature is very weak or absent. Together, the Hα, IRAC 8μm, and 11.3μm PAH images show that this structure may be a physical manifestation of a jet, streaming, and/or the local ionization, and not due to differential extinction. The IRS map in [NeII] (Figure 11b) shows a much more limited extent to the southern miniarm, indicating that to the south the ionizing radiation decreases sharply; its limiting extent can also be seen where the PAH emission suddenly weakens. The curved [NeII] structure also resembles that seen in the Hα and IRAC 8μm images. The [NeII] map also shows a stub of emission in the north corresponding to the relativistic radio jet and the origin of the northern miniarm, but it has no significant spatial extent. The [NeIII] line was mapped by IRS but with almost three times lower spatial resolution than for [NeII]. In principle the ratio of these two Ne lines would be valuable tracers of the range of the hard ionizing radiation, but the [NeIII] only shows a bright and slightly extended central peak, with only uniform, diffuse emission elsewhere.

The IRS high-res maps obtained complete spectra of a region 20.3"x13.6" (9x6 pixels) centered on the nucleus. From the full high-resolution spectrum of this large area we subtracted that of the inner region 11.3"x11.3" (5x5 pixels) in the nucleus to obtain a spectrum for the gas and PAH in the outer, rectangular zone, whose area is $3.5 \times 10^{-9}$ sr. Table 2 (column 5) lists the flux values in that outer rectangular annulus around the nucleus. The comparatively higher ratio [NeIII]/[NeII] = 1.3 (it is 0.66 around the nucleus), and the higher ratio [OIV]/[NeII] = 0.34 (compared with the inner region value of 0.10, suggest a higher level of ionization and/or somewhat higher excitation temperature in the outer region. At the edge of an irradiated cloud it is known that the [NeIII]/[NeII] flux ratio is roughly proportional to U, the ratio of the ionizing photon density to the hydrogen density (e.g., Voit, 1992). Lower densities in the outer volume ($>\sim n_H = 10^4$ cm$^{-3}$) can also enhance the ratio, and it may be that a combination of these is present in this outer zone.

The $H_2$ emission map (Figure 11c) in the 17μm 0-0 S(1) line was obtained with the IRS Long-Low module covering the full region, and shows a remarkable "S-shaped" loop emission structure that begins along the northern and southern miniarms. The long slit length of Long-Low, 168", provides coverage extending well beyond the central region in this one dimension, covering more of the curved $H_2$ structure. The map reveals that the NNW arc of $H_2$ emission extends out at least 50" from the nucleus, with some knots of emission also seen beyond. The structure is coincident (to within our ability to determine it) to the long outer extent of the northern miniarm seen in the IRAC images. The image shows that the 0-0 S(1) emission has a peak offset from the nucleus by about 20" (360pc) to the north, with a weaker peak to the SE. The excitation parameters of the off-nuclear excitation could be different from that in the nuclear region (Figure 5), but as noted earlier, the low resolution module does not have the sensitivity to detect the weaker 0-0 S(2) line in the map, and unfortunately the high resolution full coverage was limited to the 13"x13" region.

The IRS short-low image of the dust continuum emitting region between 9 and 12 μm (excluding contributions in this band from PAHs) is shown in Figure 11d. As discussed earlier, the emission contains components not only from warm silicate dust, but also a small contribution (about 30%) from stellar photospheres (e.g., Figure 7). We see no apparent deviations from a spherical distribution in the image. This is consistent with an hypothesis of an AGB wind origin for the dust (Section 5.2), as we would expect these stars to be spherically distributed around the M81 nucleus. The dust is presumably not of uniform temperature, but unfortunately the temperature we obtain from the models is a sensitive function of the 18μm /10μm ratio but not sensitive to the 10μm feature alone. We are therefore unable to say more than that a component of even hotter dust, as for example might arise from the inner edge of a torus, cannot be excluded although it is a minor contributor to the net flux.

## 7. CONCLUSIONS

We have analyzed the 4-epoch IRAC photometry, MIPS photometry, and full IRS spectroscopy of the nuclear region of M81. The comparative proximity of the galaxy (~3.63Mpc) and the small *Spitzer* beam sizes meant that we were able to obtain images and spectra of the nuclear region within a ~200pc range of the core. The nuclear region of M81 has an unusual silicate emission feature which peaks at ~10.5μm and has a FWHM much larger than that of the galactic sources. The M81 mid-IR dust emission differs dramatically from that seen in most other kinds of galaxies which have either silicate in absorption or in weak emission. A few other type 1 LINERs also seem to show this unusual profile, as do a set of quasars, although the spatial resolution and signal-to-noise ratios in these sources are inferior to that of our M81 data. We have successfully modeled this strong silicate emission as most probably being due to micron-sized porous grains with a porosity of P ~0.5 ± 0.2, and a temperature of ~275 ±30K.

We find no evidence for time variability between 2003 November and 2005 October to a precision of 1%. Furthermore, we conclude that the steeply rising infrared spectral energy distribution (SED) of the inner 230pc between 8-11μm, coupled with the filter bandpasses used, can explain the apparent photometric discrepancy with the ground-based flux measured in 1999. We have measured the fluxes in ten fine structure lines; their fluxes and ratios confirm what has been previously known, namely, that the M81 nucleus is a LINER, with some features characteristic of AGN, like [OIV] emission and weak 7.7μm PAHs, and other features of transition LINERs with contaminating starburst activity, like the infrared colors and lower excitation line ratios. Spatial mapping of the [NeII] line finds that it traces the same jet/arc/spiral structures seen in the HST Hα image and in the IRAC 8μm images. The fluxes (or flux limits) in five lines of $H_2$ yield an excitation temperature derived from a formal regression fit of 825K, with a lower value of 515K if the weakest line, S(5), is excluded from the calculation. The lower temperature limit is still somewhat warmer than the typical value seen in galaxies of about 250K. A spectral map of the 17μm 0-0 S(1) line shows an S-shaped loop emission structure that traces the spiral arcs seen at other wavelengths. The similarity of the spatial features seen at various wavelengths supports the idea that they are real physical structures and not the result of differential extinction across the region.

It is not yet understood why this LINER (plus a few other types) has this distinctive kind of porous dust in its nuclear region, or why a very similar kind of dust is seen in some quasars but yet not seen in other AGNs, or why this dust is not seen in other kinds of galaxies for that matter (Köhler et al., in prep). The M81 dust seems most likely to have been produced in AGN winds, although supernovae may also have contributed, but in either case the original grains may have been destroyed by the environment (AGN/supernovae) and then reconstituted in a process that left them quite fluffy. In non-AGN galaxies the original dust may have never be so disrupted, while in more powerful cases the dust may not have been reconstituted this way. This kind of dust may be more widespread but hard to detect. In more distant galaxies where circumnuclear star formation activity falls within the same beam that measures the nucleus – and this category includes most ULIRGs, starbursts, and even transition LINERs – the SED may be dominated by the emission or absorption character of the dust in those outer regions, thereby obscuring the emission from porous dust in a zone around the nucleus. Further analyses of additional objects, and observations with higher spatial resolutions, are needed to sort out such options.

We thank Ciska Kemper-Markwick for her advice on and assistance with dust modeling, and Achim Tappe for help with CUBISM in general and IRS calibration issues in particular. We thank Nick Devereux for providing the data for the H$\alpha$ image in Figure 2, and for his helpful comments. We also thank Lei Hao, Ciska Kemper-Markwick, Ralf Siebenmorgen, Henrik Spoon, Eckhard Sturm, Doug Whittet, and Martin Haas for providing us with the observational data of 3C273, GC Sgr A*, IRAS08572+3915, NGC3998, Cygni OB2 #12, and 3CR galaxies, respectively. We thank the anonymous referee for his/her thoughtful comments, which helped to improve this paper. HAS and MLNA acknowledge partial support from NASA Grant NAG5-10654. AL, MK, and MPL are supported in part by Spitzer Theory Programs, a Herschel Theory Program, and NSF grant AST 07-07866. This work is based in part on observations made with the *Spitzer* Space Telescope, which is operated by the Jet Propulsion Laboratory, California Institute of Technology under NASA contract 1407. Support for the IRAC instrument was provided by NASA through Contract Number 960541 issued by JPL. CUBISM is a product of the SSC and the IRS team.

REFERENCES

Alsonso-Herroro, A., *et al.* 2000, ApJ, 530, 688
Armus, L. *et al.* 2007, ApJ. 656, 148
Ashby, M. N. L. *et al.* 2009 ApJ, 701, 428
Bernard-Salas et al., 2009, ApJS, 184, 230
Bietenholz , M.F., Bartel, N., and Rupen, M.P., 2000 ApJ 532, 895
Bohren, C. F. & Husman, D. R. 1983, Absorption and Scattering of Light by Small Particles (New York: Wiley)
Blum, J. & Wurm, G. ARA&A, 46, 21, 2008
Bregman, J. N., Temi, P. and Bregman, J. D. 2006, ApJ. 647, 265
Buchanan, C. L. *et al.* 2006, AJ. 132, 401
Cameron, A. & Schneck, P. 1965, Icarus, 4, 396
Dale, *et al.* 2007, ApJ, 655, 863
Davidge T. J. and Courteau, A. 1999, AJ, 117, 278
Devereux, N. A. and Shearer, A. 2007, ApJ, 671, 118
Devereux, N. A. *et al.* 1995, AJ, 110, 1115
Devereux, N. A., Ford, H. and Jacoby, G. 1997, ApJ, 481, L71
Devereux, N. A. *et al.* 2003, AJ, 125, 1226
Dorschner, J., Begemann, B., Henning, T., Jaeger, C., & Mutschke, H. 1995, A&A, 300, 503
Draine, B. T. and Salpeter, E. E. 1979, ApJ, 231, 438
Draine, B. T., & Lee, H. M. 1984, ApJ, 285, 89
Draine, B. T. 2003a, ApJ, 598, 1017
Draine, B. T. 2003b, ApJ, 598, 1026
Draine, B. T. *et al.* 2007, ApJ, 663, 866
Draine, B.T., and Li, A., 2001, ApJ, 551, 807
Draine, B. T. & Li, A. 2007, 657, 810
Dwek, E., Galliano, F. and Jones, A. P. 2007, II Nuovo Cimento B., 122, 959
Dwek, E., Galliano, F. and Jones, A. P. 2009a, in The Evolution of Dust in Extreme Astrophysical Envrionments, EAS Pub. Series, 35, 57
Dwek, E., Galliano, F. and Jones, A. P. 2009b,
Elmegreen, D. M., Chromey, F. R., Johnson, C. O., Fernandes, A. J. L. 1995, AJ, 110, 2102
Elvis, M., Marengo, M. and Karovska, M. 2002, ApJ, 567, L107
Ercolano, B. *et al.* 2007, MNRAS, 379, 1248
Freedman, W. L. 1994, BAAS, 26, 1473
Fritz, J., Franceschini, A. and Hatziminaoglou, E. 2006, MNRAS, 366, 767
Galliano, F., Dwek, E. and Chanial, P. 2008, ApJ, 672, 214
Galliano, F. et al., 2008, ApJ, 679, 310
Gillett F., & Forrest, W. 1973, ApJ, 179, 483
Grossan, B. *et al.* 2001, ApJ, 563,687
Groves, B. *et al.* 2006, MNRAS, 371, 1559
Groves, B. *et al.* 2008, MNRAS, 391, 113
Haas, M. *et al.* 2001, A&A, 367, L9
Haas, M. *et al.* 2005, A&A, 442, 39

Haas, M. *et al.* 2008, ApJ, 688, 122
Hackwell, J. Gehrz, R., & Woolf, N. 1970, Nature, 227, 822
Hao, L. *et al.* 2005, ApJ, 625, 78
Hao, L., *et al.* 2007, ApJ, 655, 77
Heckman, T 1980, A&A, 88
Henning, T., 2010, ARA&A, 48, 21
Higdon, S. J. U. *et al.* 2004, ApJS, 154, 174
Ho, L. C. *et al.* 1997, ApJS, 112, 391
Ho, L. C., Filippenko, A. V. and Sargent, W. L. W. 1993, ApJ, 417, 63
Ho, L. C., Filippenko, A. V. and Sargent, W. L. W. 1996, ApJ, 462, 183
Ho, L. C., & Keto, E. 2007, APJ, 666, 976
Huang, J.-S., Faber, S. M 2009, ApJ, 700, 183
Jura, M. and Kleinmann, S. G. 1989, ApJ, 341, 359
Kemper, F. *et al.* 2004, ApJ, 609, 826
Kennicutt, R. C. *et al.* 2004, BAAS, 36, 1442
Kleinmann, D. Gillett, F. & Wright, E. 1976, ApJ 208, 42
Levenson, N.A., *et al.* 2007ApJ, 654, 45
Laor, A. and Draine, B. 1993, ApJ 402, 441
Li, A., & Lunine, J. I. 2003, ApJ, 590, 368
Li, M. P., Shi, Q. J., & Li, A. 2008, MNRAS, 391, L49
Li, A. 2007, in ASP Conf. Ser. 373, The Central Engine of Active Galactic Nuclei, ed. L. C. Ho & J.-W. Wang (San Francisco: ASP), 561
Lutz, D. *et al.* 1998, ApJ, 505, L103
Maiolino, R., Marconi, A., & Oliva, E. 2001, A&A, 365, 37
Markoff, S. *et al.* 2008, ApJ, 681, 905
Markwick-Kemper, F. *et al.* 2007, ApJ, 668, 107
Netzer, H. *et al.* 2007, ApJ, 666, 806
Nikutta, R. Elitzur, M., & Lacy, M. 2009, ApJ, 707, 1550
O'Halloran, B., Satyapal, S. and Dudik, R. 2006, ApJ, 641, 795
Perez-Gonzalez, P. G. *et al.* 2006, ApJ, 648, 987
Piovan, L. *et al.* 2003, A&A, 468, 559
Polletta, M. *et al.* 2006, ApJ, 642, 673
Polletta, M. *et al.* 2008, ApJ, 675, 960
Rieke, G. H. and Lebofsky, M. J. 1978, ApJ, 220, 37
Rieke, G. H. & Low, F. 1975, ApJ, 199, 13
Rigopoulou, D. *et al.* 1999, AJ, 118, 2625
Roche, P. F., Aitken, D. K., Whitmore, B. 1983, MNRAS, 205, 21
Rouleau, F., & Martin, P. G. 1991, ApJ, 377, 526
Rupke, D. *et al.* 2007, ASP Conf. Series, 373, 525
Schuster, M. T., Marengo, M., Patten, B. M. 2006, SPIE, 6270, 651
Schweitzer, M. *et al.* 2006, ApJ, 649, 79
Schweitzer, M. *et al.* 2008, ApJ, 679, 101
Siebenmorgen, R. *et al.* 2005, A&A, 436, 5
Smith, J. D. T. *et al.* 2007, ApJ, 656, 770
Sirocky, M. M. *et al.* 2008, ApJ, 678, 729

Smith, J.D.T., *et al.* 2007, PASP, 119, 1133
Spinoglio, L. *et al.* 1995, ApJ, 453, 616
Spoon, H. W. W. *et al.* 2006, ApJ, 638, 759
Spoon, H.W.W. *et al.* 2007, ApJ, 654, 49
Stein, W. & Gillett, F. 1969, ApJ, 155, 197
Sturm, E. *et al.* 2000, A&A, 358, 481
Sturm, E. *et al.* 2005, ApJ, 629, L21
Sturm, E. *et al.* 2006a, ApJ, 642, 81
Sturm, E. *et al.* 2006b, ApJ, 653, L13
Sugerman, B. *et al.* 2006, Science, 313, 196
Thompson, G.D. *et al.* 2009, ApJ, 697, 182
Thornley, M. D., *et al.* 2000, ApJ, 539, 641
Tommasin, S. *et al.* 2008, ApJ, 676, 836
Tommasin, S. *et al.* 2010, ApJ, 709, 1257
Veilleux, S. *et al* 2009, ApJS, 182, 628
Voit, G. 1992, MNRAS, 258, 841
Weedman, D. *et al.* 2006, ApJ, 653, 101
Westpfahl, D. J. 1998, ApJS, 115, 203
Whittet, D. *et al.* 1997, ApJ, 490, 729
Willner, S. *et al.* 2004, ApJS, 154, 222 (Paper I)
Wu, Y. *et al.* 2009, ApJ, 701, 658
Wu, Y. *et al.* 2010, ApJ, (submitted)

Table 1

Peak wavelengths ($\lambda_{peak}$) and widths (FWHM) of the "9.7 μm" Silicate Emission/Absorption Features of Various Objects

| Object | Nature | $\lambda_{peak}$ (μm) | FWHM (μm) | Note | References |
|---|---|---|---|---|---|
| M81 nucleus | AGN | 10.5 | 4.2 | emission | this work |
| GC Sgr A★ | ISM | 9.8 | 1.8 | absorption | Kemper et al. 2004 |
| Cyg OB2 #12 | ISM | 9.8 | 2.6 | absorption | Whittet et al. 1997 |
| IRAS 08572+3915 | ULIRG | 9.8 | 2.2 | absorption | Spoon et al. 2006 |
| 3C273 | QSO | 11 | 3.0 | emission | Hao et al. 2005 |
| NGC3998 | AGN | 11 | 3.7 | emission | Sturm et al. 2005 |

Table 2
Spectral Features in the M81 Nucleus (13"x13" region)

| Line | Transition | Wavelength (Microns) | Flux *(equivalent width in μm for PAH)* ($10^{-21}$ watts-cm$^{-2}$) | Flux Exterior[1] |
|---|---|---|---|---|
| PAH | | 6.2 | ≤ *-0.01μm* | |
| | | | ≤20 | |
| H$_2$ | (0,0) S(5) | 6.910 | 11 ± 4 | |
| H$_2$ | (0,0) S(3) | 9.665 | 9 ± 0.8 | |
| PAH | | 7.7 | ≤ *-.01μm)* | |
| | | | ≤12 | |
| PAH | | 8.6 | *-0.021μm ± 0.006* | |
| | | | 18 ± 5 | |
| [SIV] | $^2P_{3/2}$-$^2P_{1/2}$ | 10.511 | 1.9 ± 0.5 | |
| PAH | | 11.25 | *-0.110μm ± 0.001* | *-0.210μm* |
| | | | 100 ± 6 | |
| H$_2$ | (0,0) S(2) | 12.28 | 3.3 ± 0.4 | 1.7 |
| PAH | | 12.7 | *-.020μm ± .003* | |
| | | | 12 ± 2 | |
| [NeII] | $^2P_{1/2}$-$^2P_{3/2}$ | 12.814 | 34.7 ± 0.7 | 4.7 |
| [NeV] | | 14.322 | <0.8 | |
| [NeIII] | $^3P_1$-$^3P_2$ | 15.551 | 23.0 ± 0.3 | 6.3 |
| PAH | | 16.5 | *-0.011μm ± .003* | *-0.042μm* |
| | | | 4.3 ± 0.9 | |
| H$_2$ | (0,0) S(1) | 17.034 | 4.8 ± 0.2 | 2.7 |
| PAH | | 17.4 | *-0.03μm ± .0015* | |
| | | | 0.9 ± 0.5 | |
| [SIII] | $^3P_2$-$^3P_1$ | 18.713 | 7.7 ± 0.3 | 2.2 |
| [NeV] | | 24.318 | <1.1 | |
| [OIV] | $^2P_{3/2}$-$^2P_{1/2}$ | 25.890 | 3.3 ± 0.4 | 1.6 |
| [FeII] | $a^6D_{7/2}$-$a^6D_{9/2}$ | 25.988 | 3.2 ± 0.4 | 1.4 |
| H$_2$ | (0,0) S(0) | 28.22 | <0.5 | |
| [SIII] | $^3P_1$-$^3P_0$ | 33.481 | 7.1 ± 0.4 | 3.8 |
| [SiII] | $^2P_{3/2}$-$^2P_{1/2}$ | 34.815 | 19.2 ± 0.4 | 10.2 |

[1] Flux exterior to the nuclear region: these fluxes were measured with the HiRes module in a rectangular ring of area 3.5x10$^{-9}$ sr that excluded the nucleus. For PAH features the emission is given both as an equivalent width and in watts-cm$^{-2}$. The outer boundary of the ring was defined by the region mapped with the hi-res modes (a rectangle of 6x9 short-high pixels of 2.3 arcsec each), and the inner boundary of the ring was a square of 5x5 pixels around the nucleus.) Flux uncertainties are the same as those for the nuclear region.

Table 3:

Parameters for Models Best Fitting the Mid-IR Dust Emission Observed in the M81 Nucleus

| Model | Note | Porosity P | Dust Size (a) | Dust Composition | $T_w$ (K) | $m_w$ ($M_\odot$) | $T_c$ (K) | $m_c$ ($M_\odot$) | $\chi^2/N$ |
|---|---|---|---|---|---|---|---|---|---|
| 1 | recommended | 0.5 | 5 µm | $MgFeSiO_4$ | 275 | 0.034 | 40 | 1500 | 1.87 |
| 2 | | 0 | 3.2 µm | $MgFeSiO_4$ | 307 | 0.029 | 40 | 2045 | 3.12 |
| 3 | | 0.3 | 3 µm | $MgFeSiO_4$ | 258 | 0.038 | 40 | 2250 | 2.01 |
| 4 | | 0.7 | 9.5 µm | $MgFeSiO_4$ | 278 | 0.041 | 40 | 1730 | 2.65 |
| 5 | | 0 | 1.2 µm | "astronomical" silicates | 207 | 0.11 | 45 | 950 | 4.57 |
| 6 | | 0.5 | 1.5 µm | "astronomical" silicates | 195 | 0.15 | 45 | 500 | 5.32 |

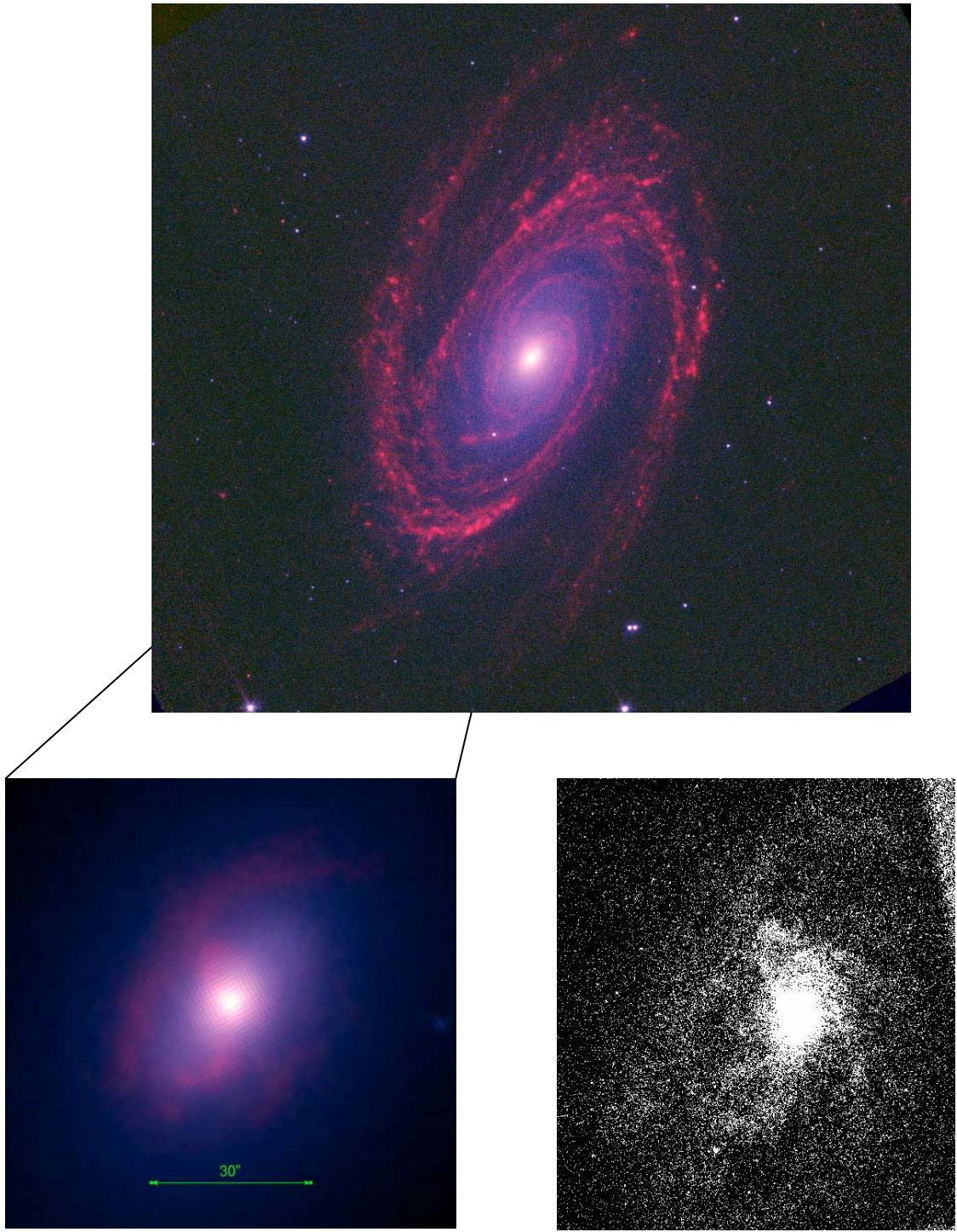

Figure 1: (a) The IRAC 3-color (3.6μm-blue; 4.5μm-green; 8.0μm-red) images of M81 (north is up and east is left; the intensity scaling is logarithmic). The stellar continuum has been subtracted as in Paper I (see text Section 2.1); (b) the M81 nuclear region (same color scheme); (c) an Hα image of the same region as in (b) (Devereux et al. 1995). Note the close correspondence between the Hα structures and those seen in IRAC (Figure 1b), and in the 11.3μm PAH image (Figure 11a).

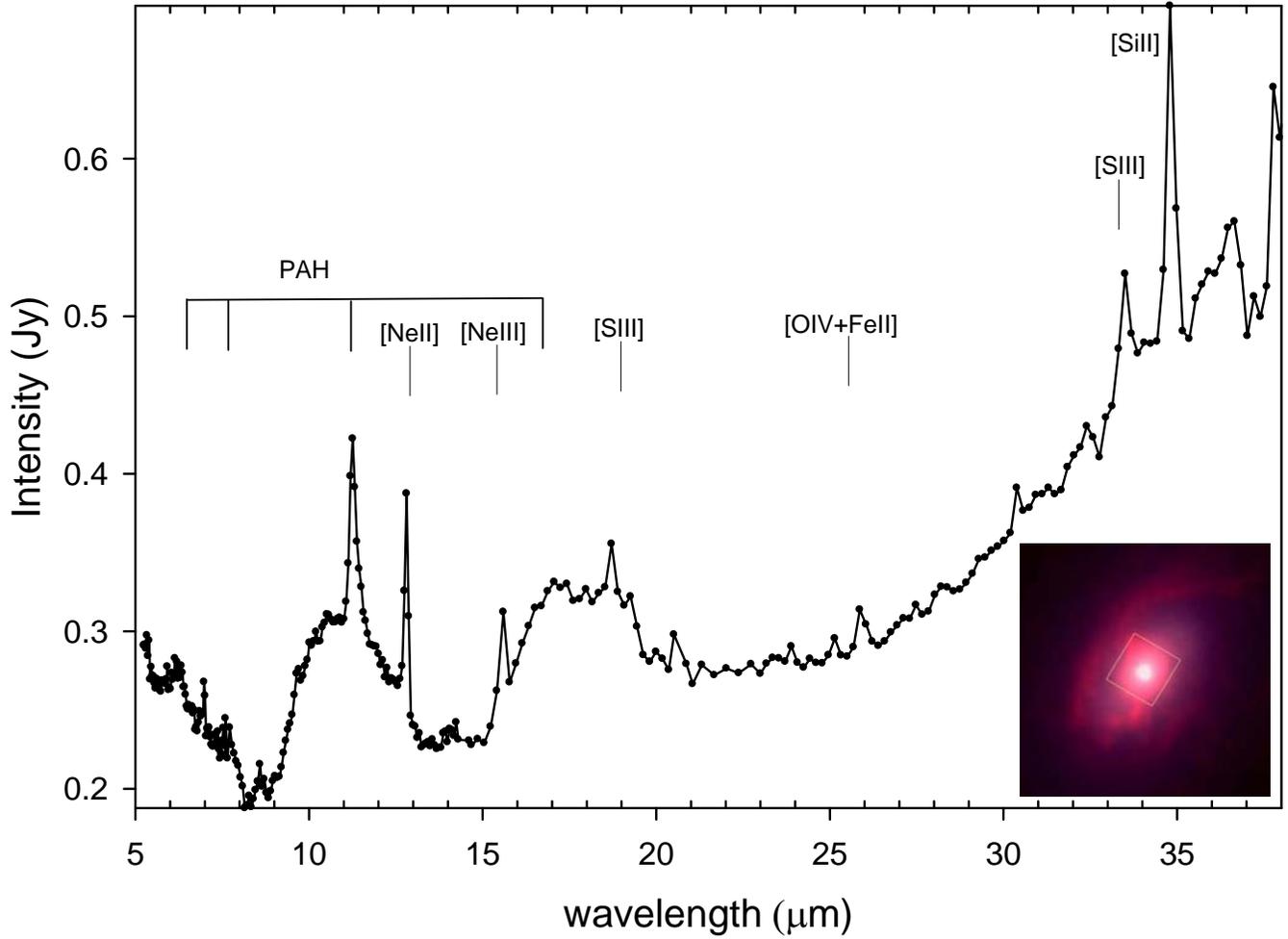

Figure 2: The IRS low resolution spectrum of the nuclear region of M81. The data were extracted from the four IRS low-res bands across the central 13.0x13.0" (shown in the boxed region in the insert, from Figure 1b). Strong lines are indicated (see also Figure 5).

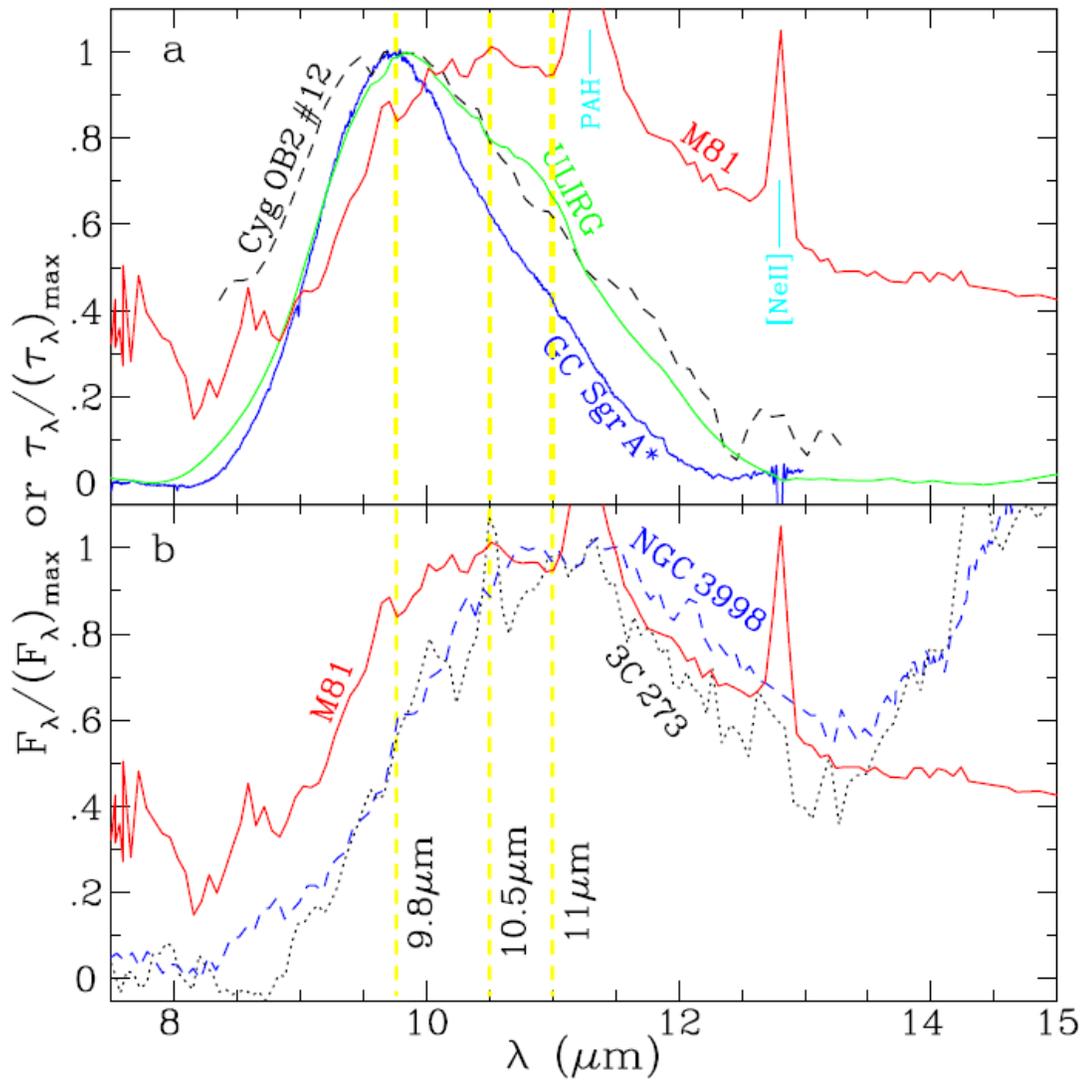

Figure 3: Comparison of the M81 nucleus mid-IR silicate emission spectrum (red solid line) with (a) the silicate absorption optical depths of the local ISM toward Cyg OB2 #12 (black dashed line), the ISM toward the Galactic Center object SgrA* (solid blue line), and IRAS08572+3915, a ULIRG (solid green line); (b) the silicate emission spectra of 3C273, a bright quasar (black dotted line), and NGC 3998, a low-luminosity AGN (blue dashed line). The yellow vertical lines mark the peak wavelengths of the silicate emission/absorption features of these objects. The sharp lines at ~10.5μm and 14.3μm in 3C273 are from [SIV] and [NeV], respectively.

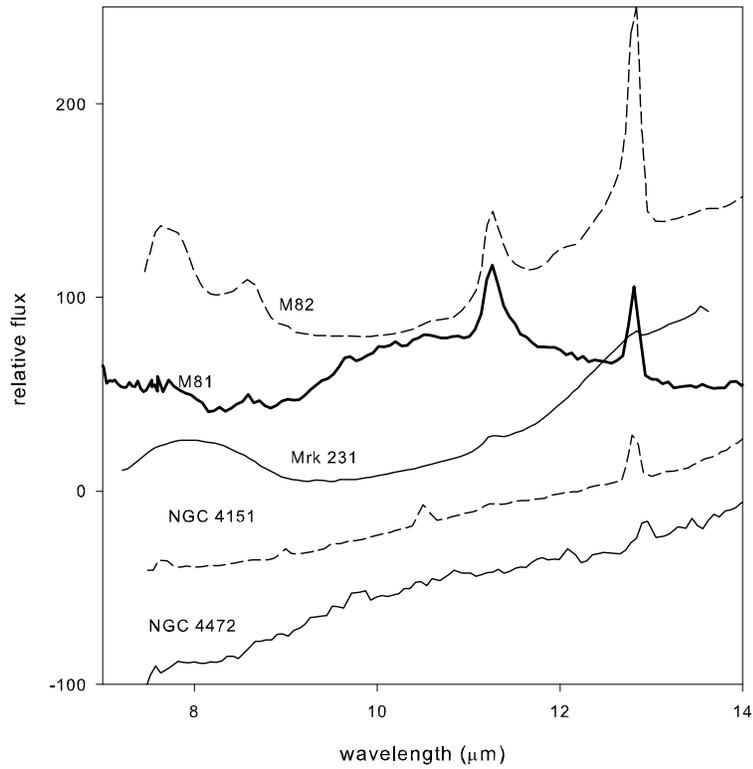

Figure 4: The IRS mid-IR dust emission spectra of four galaxies representative of types, as compared with that of the M81 nucleus (in bold): M82 (a starburst galaxy), Mrk 231 (an AGN), NGC4151 (a type1 Seyfert), and NGC4472 (an elliptical galaxy). The fluxes of each are scaled and offset for ease of comparison.

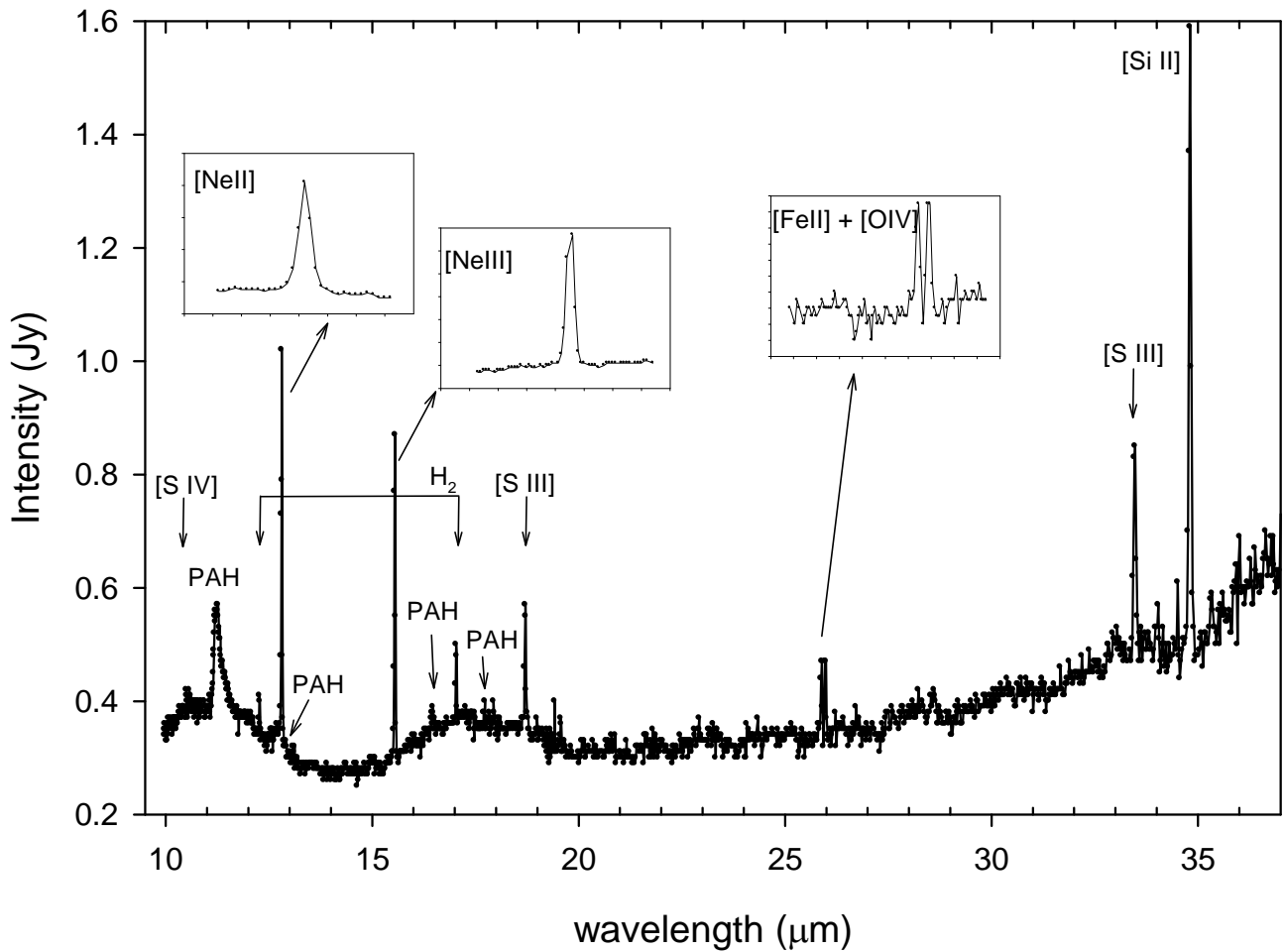

Figure 5: The IRS high-resolution spectrum of the nuclear region of M81; insets show as examples the [NeII] 12.8 μm, [NeIII] 15.55 μm, and the two adjacent [OIV] 25.89 μm and [FeII] 25.99 μm lines; the 11.3, 12.7, 16.5μm and 17.4μm PAH features, $H_2$ lines, and fine structure lines are also indicated (also see Table 2).

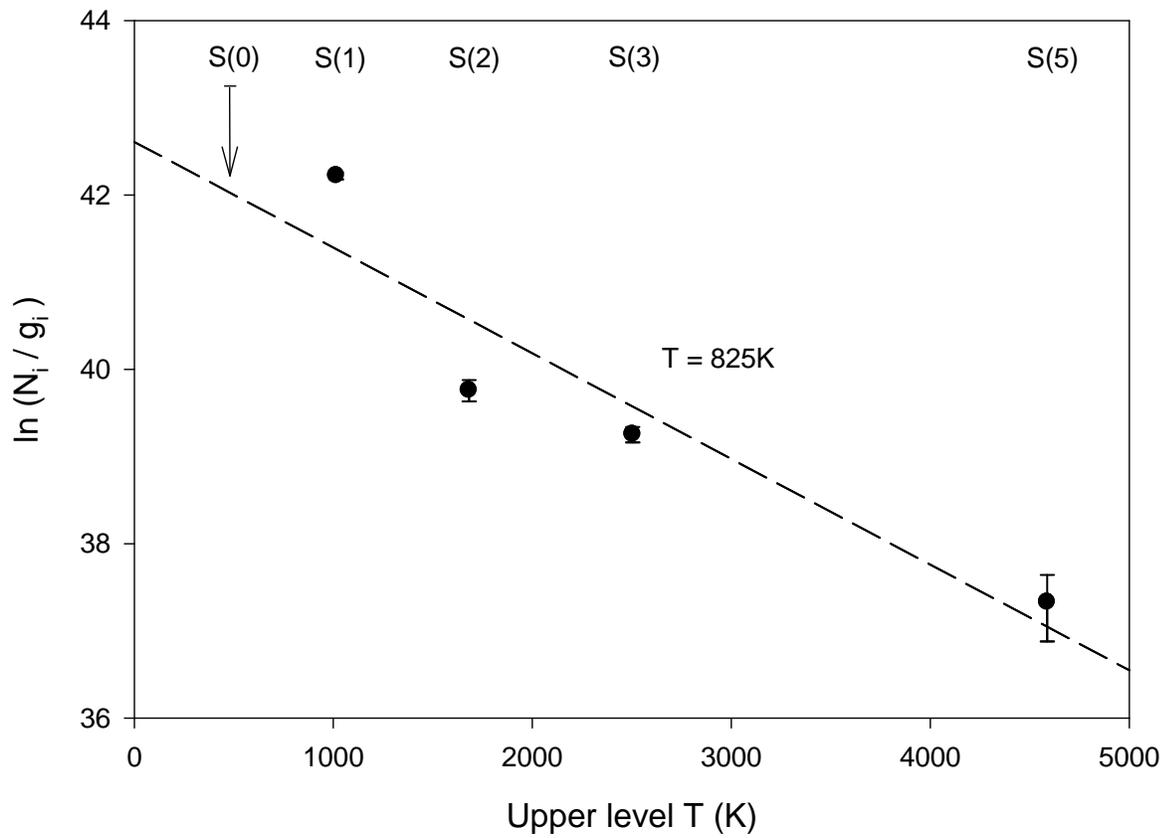

Figure 6: The $H_2$ excitation in the nucleus of M81. The dotted line regression to the four detections corresponds to a temperature of about 825K; excluding the noisier S(5) line, the value is T=515K, with the range of possible slopes giving temperature values between 270K and about 1000K. ($N_i/g_i$ is the population of level i normalized by its degeneracy; see text.)

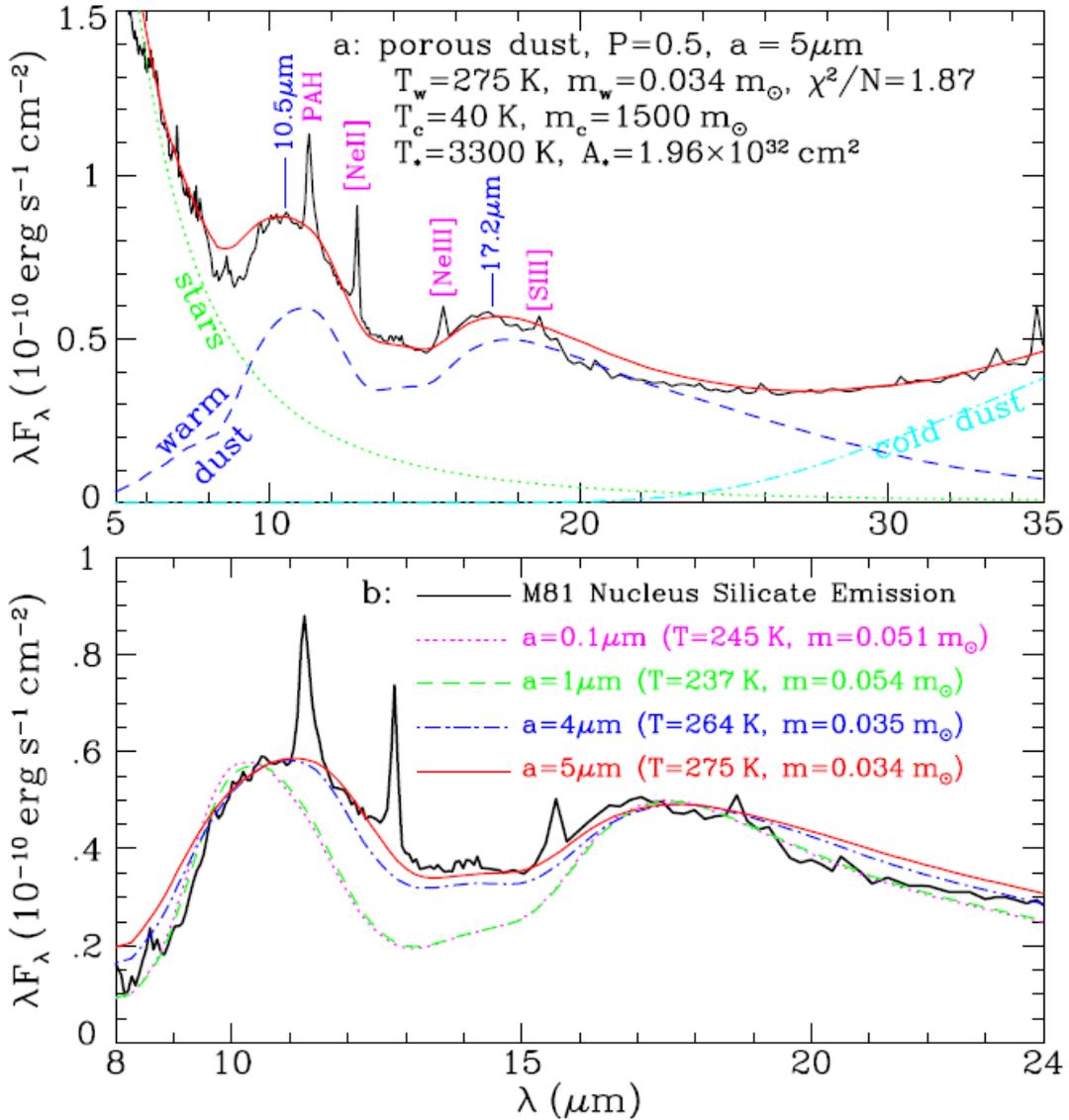

Figure 7. Fits to the 5-35μm IRS silicate dust emission in the M81 nucleus: Top panel (a): Model fit (red solid line) to the M81 nucleus (black solid line). The model consists of a population of warm dust ($T_w \approx 275$K, $m_w \approx 0.034$ $M_\odot$ ; blue dashed line) and cold dust ($T_w \approx 40$K, $m_w \approx 1500$ $M_\odot$ cyan dot-dashed line), as well as a stellar continuum (green dotted line). The dust is a porous mixture of amorphous olivine $MgFeSiO_4$ (Dorschner et al. 1995)and amorphous carbon, with a porosity of P=0.5 and a size of a=5 μm . Bottom panel (b): Same as (a) but now with the stellar continuum subtracted, and illustrating the differences between Si-O (peaking at ~10.5μm and O-Si-O (which peaks at ~17.2μm) silicate emission bands (obtained by subtracting the stellar continuum and the cold dust component from the IRS spectrum) and the effects of dust size a. For grains of a<1μm (and P=0.5), the model profiles are insensitive to size a, since they are in the Rayleigh scheme at >8μm .

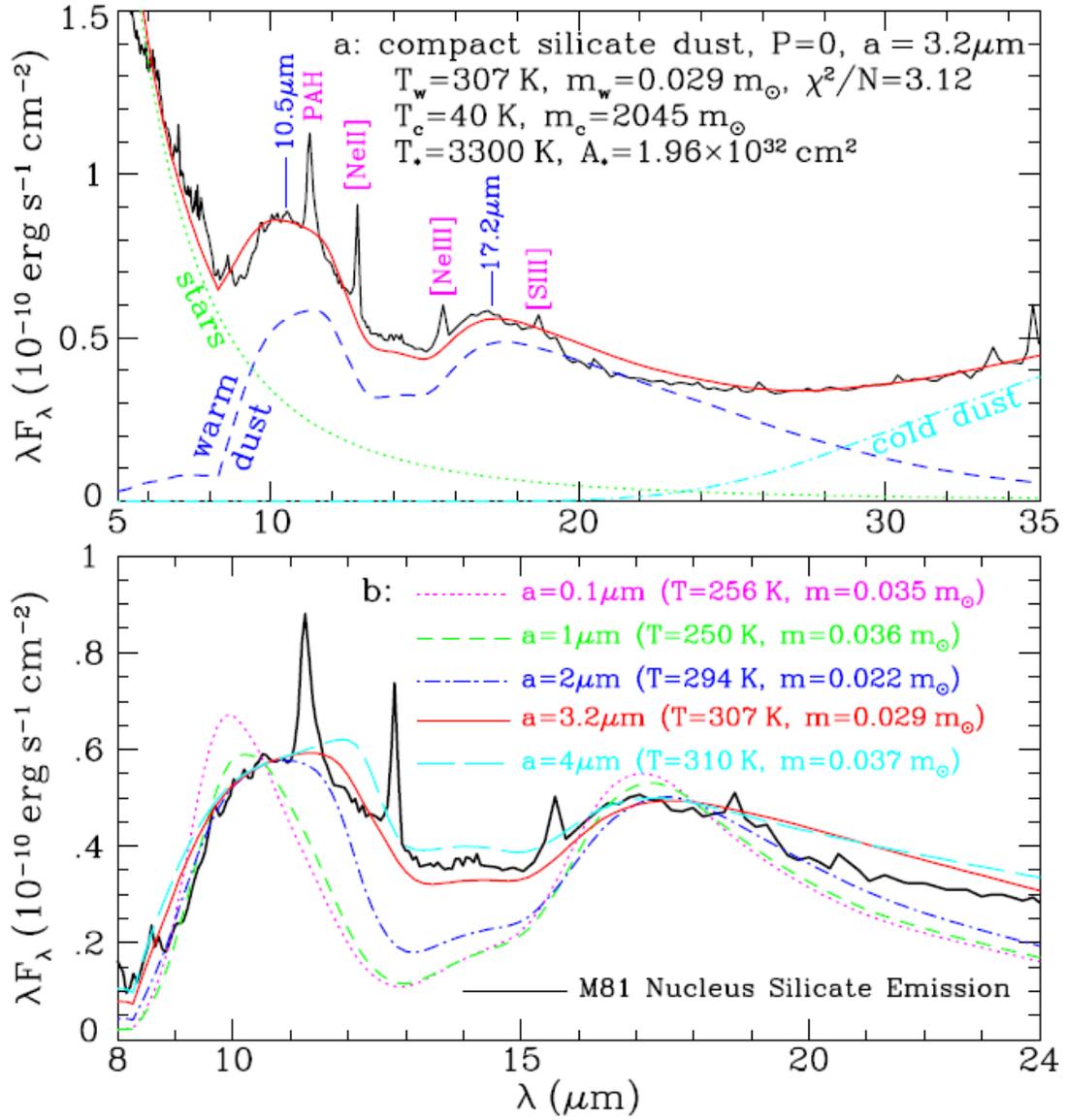

Figure 8: Same as Figure 7, but for compact, pure silicate dust. Very large compact dust grains would be required to produce a reasonable fit to the observations.

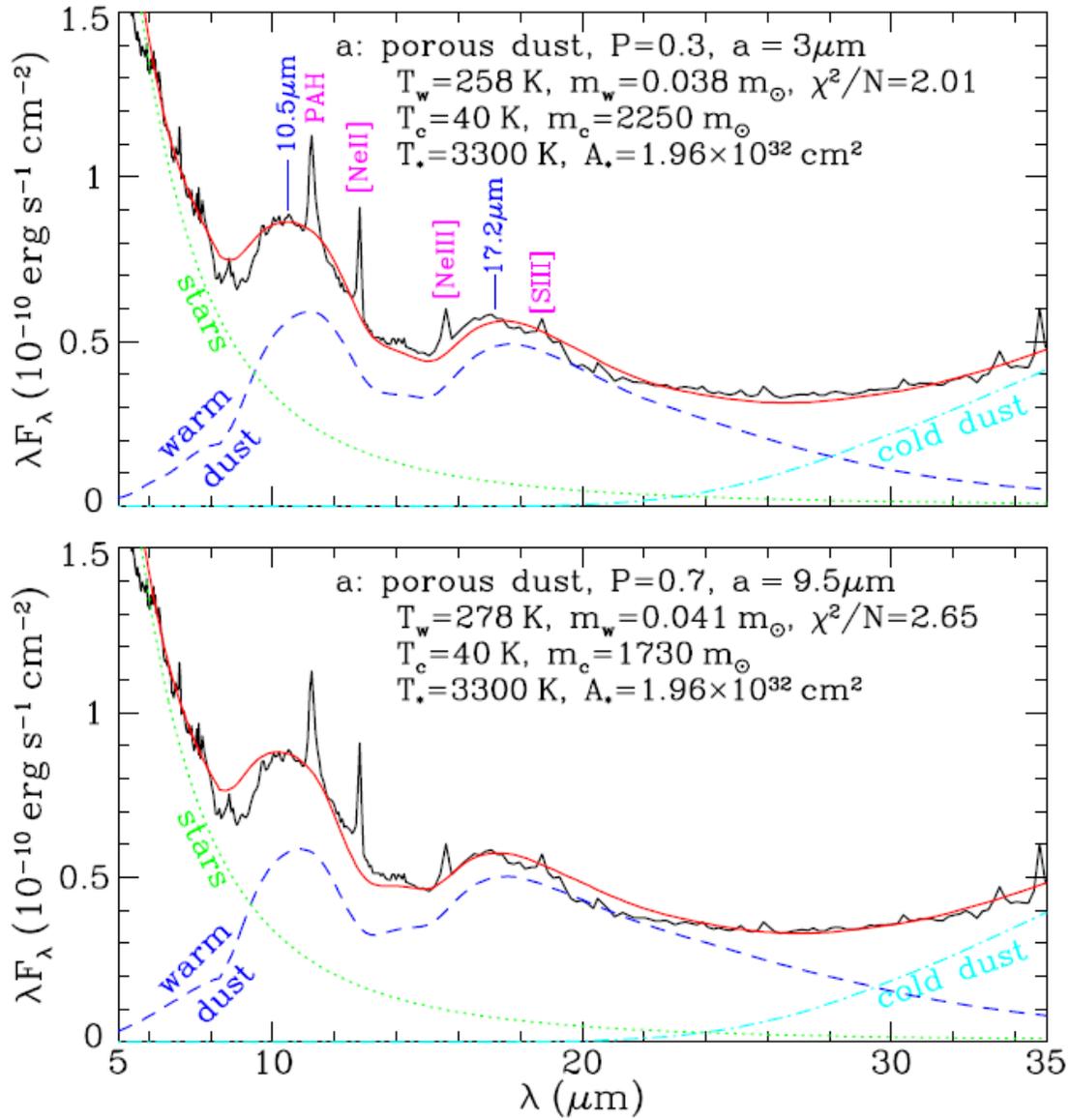

Figure 9: Illustrating the effects of varying the dust porosity. Same as Figure 7a, but for the case with (a) dust with P=0.3 and a=3μm, and (b) dust with P=0.7 and a=9.5μm .

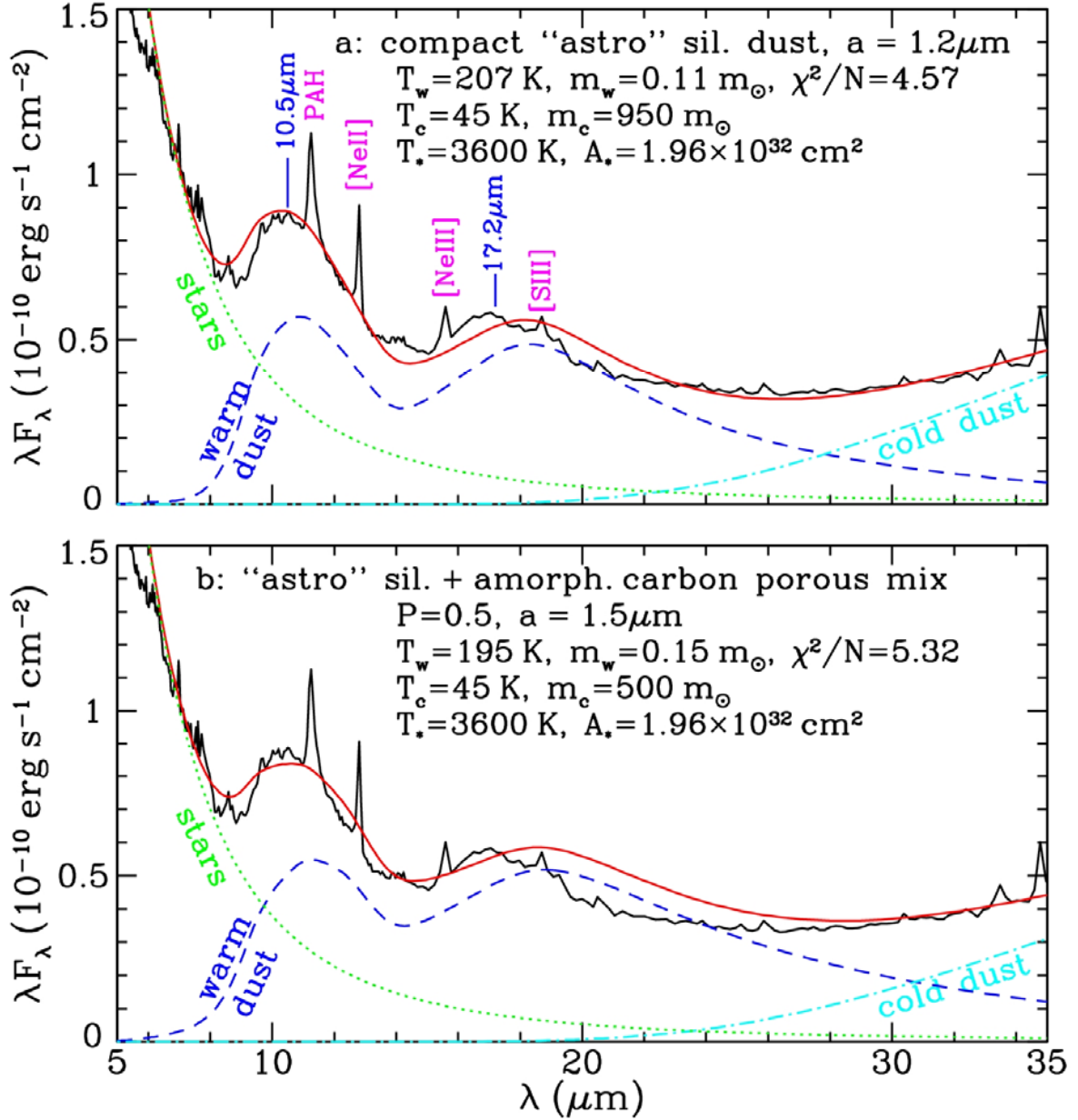

Figure 10: Comparing best fits of "astronomical silicates" to porous dust. Top panel (a): Fitting the IRS spectrum of M81 (solid black line) with COMPACT dust (solid red line) consisting of a warm population with $T_w \approx 207K$ and $m_w \approx 0.11$ Mo (dashed blue line) and a cold population with $T_c \approx 45K$ and $m_c \approx 950$Mo (dot-dashed cyan line). The continuum emission at $\lambda<8\mu m$ is approximated by stellar photospheric emission from M0III stars of $T_* \approx 3600K$ (dotted green line). The silicate composition is taken to be that of the Draine-Lee (1984) ``astronomical'' silicates. Bottom panel (b): Same as (a) but for POROUS silicate dust of $a = 1.5\mu m$ (parameters as specified in the labels).

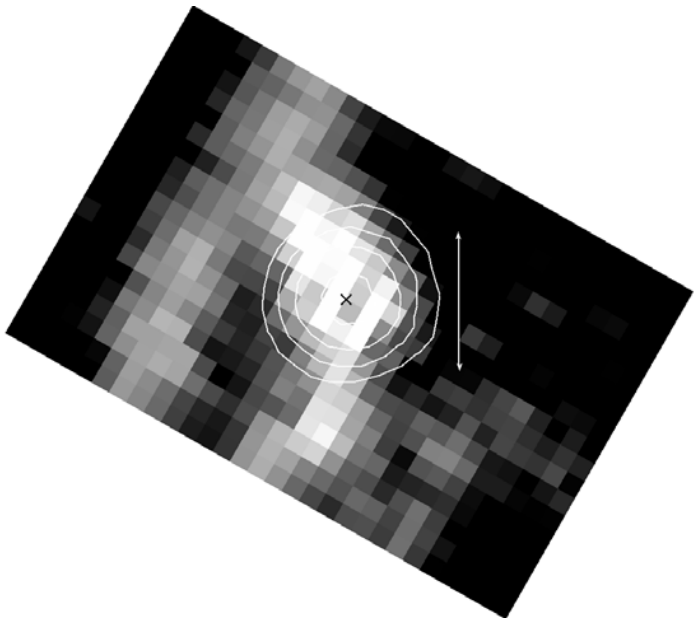

(a)

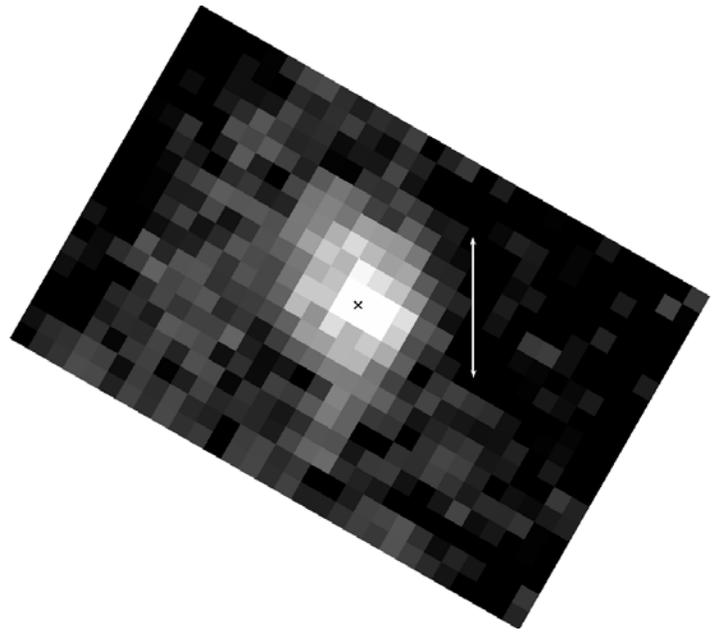

(b)

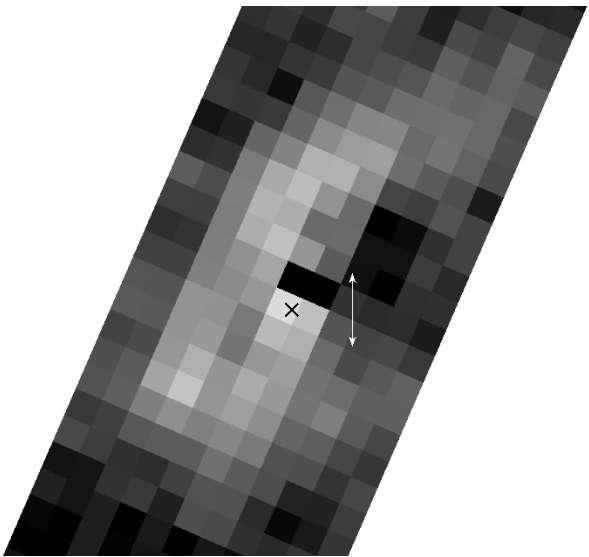

(c)

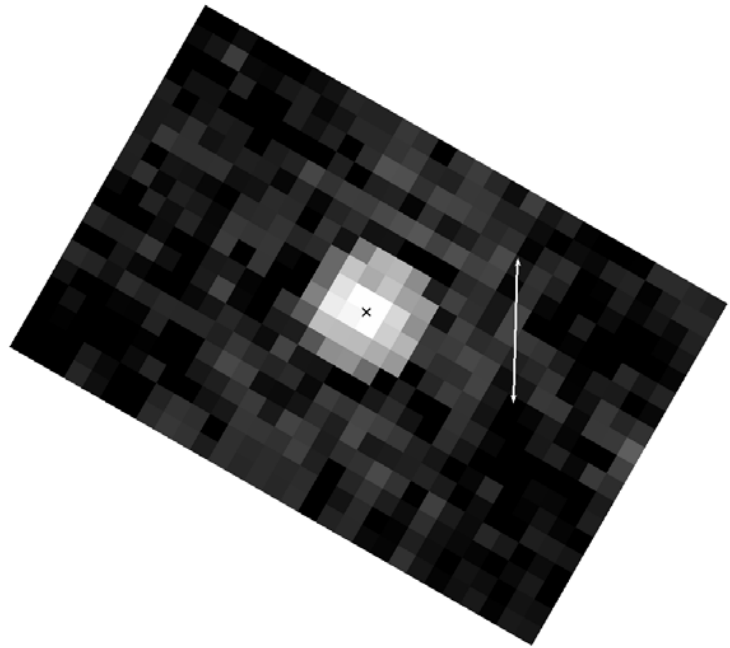

(d)

Figure 11: Spectral images of key features in the M81 nucleus. Each figure is centered on the M81 nucleus as marked by an X (north is up), and the distance scale is shown by a bar 13.0" in length in all images. The intensities are displayed using logarithmic scaling in all cases. (a) PAH 11.3μm emission, background subtracted, from the IRS-short-high (SH) model. The contours show the 10.5μm silicate dust distribution (taken from Figure 11d), with the levels at 3%, 6%, 13%, and 25% of the peak. (b) [NeII] 12.8μm emission, background subtracted, from the IRS short-high model; (c) $H_2$ 17μm background subtracted, from the IRS long-low model. Because IRS-LL mapped a larger region than the SH modules, this image covers 100" in height; (d) 9-12μm dust continuum emission feature as imaged by the IRS short-low module. The dust emission map excludes contributions from the PAH feature at 11.3μm; contributions from stellar photospheric emission are not excluded, and are estimated to be about 35% of the total (the cold dust contribution is negligible at this wavelength). The dust continuum emission, unlike the spectral line emissions, shows no sign of spatial asymmetry or extent.